\documentclass[12pt,titlepage]{article}

\usepackage[margin=1.0in]{geometry}

\usepackage{setspace} % changed by BJ on 2-21-13! remove when on ipad

\usepackage{amsmath}

\usepackage{subfigure} %remove when on ipad

\usepackage{graphicx}

\usepackage[round,numbers,sort&compress]{natbib}
\title{Analysis of kinesin mechanochemistry via simulated annealing}

\author{B. D. Jacobson, L. J. Herskowitz, S. J. Koch, S. R. Atlas\thanks{
           Corresponding author. Email: susie@sapphire.phys.unm.edu. Address:
           Department of Physics and Astronomy,
	  University of New Mexico,
	   Albuquerque, NM~87131, U.S.A.,
	   Tel.:~(505) 277-2616, Fax:~(505) 277-1520}\\
	 Department of Physics and Astronomy \\
	University of New Mexico, Albuquerque, NM 87131}

\date{}

\begin{document}

% generate the title page from the info in the headers above
\maketitle

% 200 words max Abstract
\abstract{The molecular motor protein kinesin plays a key role in fundamental cellular processes such as intracellular transport, mitotic spindle formation, and cytokinesis, with important implications for neurodegenerative and cancer disease pathways.  Recently, kinesin has been studied as a paradigm for the tailored design of nano-bio sensor and other nanoscale systems.  As it processes along a microtubule within the cell, kinesin undergoes a cycle of chemical state and physical conformation transitions that enable it to take $\sim$ 100 regular 8.2-nm steps before ending its processive walk. Despite an extensive body of experimental and theoretical work, a unified microscopic model of kinesin mechanochemistry does not yet exist.  Here we present a methodology that optimizes a kinetic model for kinesin constructed with a minimum of \textit{a priori} assumptions about the underlying processive mechanism. Kinetic models are preferred for numerical calculations since information about the kinesin stepping mechanism at all levels, from the atomic to the microscopic scale, is fully contained within the particular states of the cycle: how states transition, and the rate constants associated with each transition. We combine Markov chain calculations and simulated annealing optimization to determine the rate constants that best fit experimental data on kinesin speed and processivity.

\emph{Keywords:} Molecular machines; Kinesin; Molecular motor proteins; Nanoscale biophysics; Kinetic modeling; Kinetic Monte Carlo;
ATP hydrolysis; Markov theory; Numerical optimization; Simulated annealing}

% New page
\clearpage

\section*{\small{INTRODUCTION}}

Kinesin is a molecular motor protein that processes hand-over-hand on microtubules, at each of its 8 nm steps hydrolysing a single molecule of ATP \citep{Coy1999,Yildiz2004}. Hydrolysis occurs at one of the two catalytic domains (heads) responsible for procession \citep{Hackney1994}. These interact through the neck linker, a polymer of 10-16 amino acids in length connecting the heads. The walk is initiated by ATP binding to one of the heads, causing the neck linker to dock \citep{Asenjo2006,Brunnbauer2012,Duselder2012,Miyazono2010,Rice2003}. The resulting strain on the second head biases its diffusional search toward the next binding site on the microtubule \citep{Rice1999,Carter2005,Yildiz2008}.

Each step in kinesin procession can be modeled as a cycle composed of stages in which kinesin assumes different chemical states and physical conformations \citep{Moyer1998,Cross2004,Clancy2011a}. This cycle is the starting point in developing a kinetic model for kinesin, in which all information about the kinesin step is contained in the stages of the cycle, in how they transition from one another, and in the rate constants associated with each transition. For this ability to embody information on the mechanochemical gating of the kinesin step and yet being relatively simple, kinetic models are the usual choice for performing numerical simulations of the kinesin walk on a microtubule \citep{Hyeon2011,Liepelt2007,Liepelt2009,Liepelt2010}. However, the small length scales and short time scales prevent experiments from resolving each stage of the kinesin step. Therefore the catalytic cycles and resulting kinetic models, must be deduced from what is known about ATP hydrolysis, interaction of kinesin heads with the hydrolysis products, and neck linker physical characteristics.
%A search of recent literature on the topic shows much advance in this kind of modeling, but there is still little consensus on one generally accepted kinesin catalytic cycle. Experimental data on some rate constants are also controversial: For instance, according to several recent papers,the experimentally observed rate of inorganic phosphate release from the catalytic domain is too low to account for hydrolysis occurring at a rate of 1 ATP/step. All the uncertainties involved with modeling the kinesin step fail to provide a clear theoretical picture of the states that constitute a kinesin step. In our efforts to make theoretical progress in kinesin step modeling, we have developed an optimization routine for catalytic cycles that searches in the parameter space of rate constants for a set that better reproduces the experimental observations on kinesin speed and processivity at varying concentration of ATP and hydrolysis products. This optimization scheme may serve as a guide to choose the cycle that is most plausible to realize the kinesin step.

Our numerical investigation of the kinesin walk aims to reproduce the accepted experimental results for kinesin velocity and run length as a function of ATP concentration and at different experimental conditions with respect to the concentration of ADP and inorganic phosphate (P$_i$).

The goal of performing an optimization routine involving simulated annealing to fit the experimental data for kinesin speed and processivity is two-fold: firstly, we aim to model the kinesin cycle in a unified way, including all states allowed by experiments. This way we can identify the states that are visited often during the walk (common states) and those which are visited rarely, but may have an impact on the overall performance of the protein walk. Secondly, we can find ``better" rate constants: at the end of the simulation, some rates will be very close to the values found experimentally, but others will be off. This means that either they are not contributing to the kinesin walk, or they need to be better investigated by experiments and theoretical models.

%{\it Mention here arguments for building the model as we did and arguments for choosing the rate constants.}

%Citations are made using the \emph{citep} command, e.g., one paper
%\citep{el-Kareh_etal93} or more papers
%\citep{el-Kareh_etal93,Chen_Nicholson00}.  It works fine with a single
%author, two authors, or more.  Books are cited in the same way
%\citep{Callaghan91}, book chapters as well \citep{Stiles_Bartol01},
%including a chapter in press \citep{Stiles_etal04}.  Abstracts can be
%handled too \citep{Tao_etal02}.  Pages can be included in a citation
%\citep[pp.~12--18]{Callaghan91}.  Citations in a text form (i.e., author
%name followed by the usual reference number in parenthesis) can be
%done using the \emph{citet} command, e.g., ``an approach used
%by \citet{el-Kareh_etal93}''.

\section*{\small{METHODS}}

%\citep{Block07}

%Text can reference Eq.~\ref{eqn:symmetry}
%\begin{equation} \label{eqn:symmetry}
%   \Phi(\vec{r}) = \Phi(-\vec{r})
%\end{equation}
%anywhere, as long as it is numbered.

%\subsection*{Theory}

%It is currently well-accepted through extensive experimental and theoretical evidence that kinesin hydrolyses one ATP per 8.2-nm step \citep{Coy1999,Yildiz2004}. However, there is currently only a partial picture of the sequence of chemical and mechanical transitions that begins with hydrolysis and ends with processive motion.

%In principle, in order to determine the dynamical evolution of the protein, molecular dynamics (MD) simulations can be performed. MD has the advantage of providing a detailed picture, at the atomic scale, of all mechanochemical configurations assumed by the protein at all times over the duration of a step. However, since a single kinesin step occurs on the timescale of 1$\mu$s \citep{Zhang2012} and simulation timesteps are on the order of 0.1-1.0 fs \citep{Li2012}, it is not yet possible to simulate the dynamics of the protein so as to capture a complete processive series for direct comparison with experiment. On the other hand, the study of the system's dynamics can be greatly simplified if the stochastic nature of the process is considered, and the mechanochemical configurations are mapped onto discretized states.

\subsection*{Mechanochemical-state model}

%the first step is to decide what states you've got (discuss this before starting)

The first step towards finding a kinetic model for kinesin is choosing which states will be present in the model rate constants and transitions between states follow from this choice. We aim to be as general as possible, utilizing only experimental data to constrain our model.

We start by listing all states that could be included in kinesin mechanochemistry, which can be found by combinatorics: there are 4 different chemical states for each of the two catalytic cores, which can be empty (APO state, also known in the literature as the rigor state \citep{Goulet2014}) or bound to either ATP, or ADP, or ADP+P$_i$; and two mechanical conformations of the each head with respect to the microtubule (bound or unbound to the MT). Combining all these possibilities and assuming that the heads are distinct, we obtain 80 possible states. If the catalytic cores are indistinguishable in the stochastic simulation, the number of states is reduced to 48. This number can be further reduced by incorporating the following assumptions based on experimental evidence: (i) kinesin fully detaches from the microtubule and ends its processive run in the ADP-ADP state %making the ADP-ADP state the only one to appear at the level of 0HB states
\citep{Yajima2002}; (ii) the neck linker is in the docked configuration on any head containing ATP or ADP+P$_i$, and is undocked otherwise \citep{Rice1999,Sindelar2007,Kikkawa2006}; and (iii) if both heads are bound to the microtubule, the neck linker cannot be in the docked configuration on the front head due to strain \citep{Guydosh2009}. In the third assumption, even though \citep{Guydosh2009} also forbids a two-head bound state with the neck linker undocked in both heads, those states were added due to evidence that some of these states may appear in mechanisms that include backstepping and futile hydrolysis \citep{Clancy2011a}. We are then left with a 25-state model that embodies all possible chemical bonding states in the system, taking into account the three essential features of the kinesin catalytic core during the cycle: nucleotide bonding, neck-linker configuration, and microtubule binding. This 25-state model includes all states commonly proposed in the literature for the kinesin catalytic cycle (see, for instance Refs. \citep{Schief2001,Liepelt2010,Hyeon2011,Clancy2011}).

In our stochastic chemical kinetics scheme, the set of discrete states to be used in the model is first enumerated as shown in Fig.1. These states correspond to different nucleotide occupations in each catalytic core, distinct neck linker conformations (docked or undocked to each head) and the various binding states with respect to to the microtubule (one-head, two-head or detached).
In Fig. 1(a), we have all the allowed two-head-bound (2HB) states, and in Fig. 1(b), we picture the allowed one-head-bound (1HB) states. We identify as chemical transitions all those indicated by the arrows in Fig. 1. In order to distinguish the mechanical transitions, we double the 8 allowed 2HB states offset by two rows and set Fig. 1(a) below Fig. 1(b), in such way that the 16 states are perfectly stacked, as shown in Fig. 2, left. Figure 2 also includes the detached state (0HB), on the top layer. Labels corresponding to the unique states are shown in Fig. 2, right. Once the arrays of chemical transitions are disposed in layers, we can characterize the mechanical transitions as the ones that cause a change in level (attachment/detachment of one head to/from the microtubule). Such transitions are indicated by the arrows in Fig. 3. In summary, Figs. 1-3 show all possible transitions between the 25 allowed states, where transitions along the ``$x$''- direction are chemical (within a level) and transitions along the ``$y$''- direction are mechanical (inter-level).

The transitions to each state, represented by the arrows in Figs. 1-3, are associated to a set of rate constants $\{k_{i\rightarrow j}\}$, related to the probabilities of transitioning from state $i$ to state $j$. The rate constants we use to initialize the problem are found in Ref. \citep{Herskowitz2010} and references therein (see Table 1, column ``Initial''). Figure 4 shows the matrix form of the rate constants, where the states $i, j$ correspond to the header column and row, respectively, by the numerical labeling shown in Fig. 2. A matrix of rate transitions $\omega_{i\rightarrow j}$ can be derived from the rate constants matrix of Fig. 4. We use the relation $\omega_{i\rightarrow j} = k_{i\rightarrow j} I_{i\rightarrow j}\{[X]\}$ from \citep{Liepelt2007}, where the indicator function $I_{i\rightarrow j} \{[X]\}$, is unity for every transition that does not involve the capture of ATP, ADP or P$_i$ by any of the catalytic heads, and is equal to $[X]$, if the chemical reaction results in attachment of $X$, where $X$ can be ADP, ATP or P$_i$. Combined, the set of 25 discrete states and matrix of rate transitions based on Fig. 2 contain all information necessary for a kinetic simulation of the kinesin stepping mechanism.

\subsection*{Kinetic Monte Carlo}

The sequence of states assumed by kinesin is assumed to be Markovian, meaning that the system carries no memory of states previously occupied, and moving to the next state in the process depends only on the current state of the system.

For a Poisson process the time to transition from one state to the other has exponentially decaying statistics and the probability density function of leaving a state $i$ for $t > 0$ is:

\begin{equation}
p_i(t;a_i) = a_i e^{-a_it},
\end{equation}
where $a_i = \sum_{j} \omega_{i\rightarrow j}$. The total time needed to leave a state $i$ is \citep{Schwartz2008}:

\begin{equation}
\tau_i = \int_0^\infty t\:dt\: p_i(t;a_i) = \frac{1}{a_i}.
\end{equation}

We use the property of exponentially decaying transition probability density functions for the time to escape a state in kinetic Monte Carlo and Markov theory to estimate the mean run time of the protein and therefore describe the kinesin stepping mechanism.

Kinetic Monte Carlo (KMC) is a stochastic algorithm developed to study the dynamics of chemical reactions \citep{Gillespie2007,Sumathy2012}. The algorithm consists in drawing two random numbers $r_1, r_2$ in [0,1] before each state transition. The first random number gives us the time to transition
 $\Delta t_{i\rightarrow j}$ from state $i$ to state $j$ according to the rule:

\begin{equation}
\Delta t_{i \rightarrow j} = -\frac{\ln(r_1)}{a_i},
\end{equation}
with $a_i$ as defined previously. The time to transition is therefore drawn from an exponentially-decaying distribution. The state $j$ to which the system will transition after a time $\Delta t_{i\rightarrow j}$ is the first state for which the following inequality is true:

\begin{equation}
a_i^{(j-1)} < a_i r_2 \le a_i^{(j)},
\end{equation}
with $a_i^{(j)} = \sum_{l=1}^j \omega_{i\rightarrow l}$.  The simulation runs until the detached state (labeled 25 in Fig. 2) is reached. The advantages and disadvantages of kinetic Monte Carlo stem from the very simple and repetitive nature of the method: whilst we can generate a long chain of states and therefore obtain a detailed description of which transitions occur in each run, meaningful observables such as run length and run time can only be extracted by averaging over many runs, making the method computationally expensive to be used in the optimization.

\subsection*{Markov Chain}

The Markovian nature assumed for the dynamical process grants us the use of Markov theory to determine the observables of the system \citep{Gillespie2007,Clancy2011a,Kutys2010,Hughes2011}. The 25-state kinetic model is discrete, but states can transition at any time $t$. Therefore, our model corresponds to a continuous-time Markov chain (CTMC), where the time spent in each state is associated to an exponentially decaying probability, as given in Eq. 2. To determine the number of times a state is visited before reaching the detached state, the homogeneous CTMC can be reduced to an embedded Markov chain, treated as a discrete-time Markov chain (DTMC). One major characteristic of the 25-state model is that every state can be visited at a time $t$ until reaching the detached state, when the walk stops. In Markov theory, the states that can transition to another are called \emph{transient} states, and the states that once attained, cannot transition to any other state are known as \emph{absorbing} states. For the 25-state kinetic model of Fig. 2, the absorbing state is the fully detached ADP-ADP state, and all others are transient states. Therefore, our embedded Markov chain is an absorbing Markov chain.

The matrix elements of an absorbing Markov chain are the transition probabilities to go from state $i$ to state $j$, defined as $p_{ij} = \omega_{i\rightarrow j}/\sum_{i\ne l}\omega_{i \rightarrow l}$, for $i\ne j$, and $p_{ii} = 0$, except for the absorbing state, where $p_{ii} = 1$. The transition probability matrix is of the form:

\begin{equation}
P = \left(
\begin{array}{cc}
Q & R \\
0 & I
\end{array}\right),
\end{equation}
where $Q$ is the transition probability matrix between transient states, $R$ is the transition probability matrix to go from the transient states to the absorbing states, $I$ is the identity matrix, and 0 is the zero matrix. From the transition probability matrix $Q$, we can determine the number of times $f_{ij}$ that state $j$ appears if we start the cycle at state $i$ (using Einstein summation rule):

%\begin{equation}
%Q = \left(
%\begin{array}{ccc}
%0 & q_{12} & \cdots \\
%q_{21} & 0 & \cdots \\
%\vdots &\vdots & \ddots
%\end{array}\right),
%\end{equation}

\begin{equation}
f_{ij} = q_{ij} + q_{ik}q_{kj} + q_{ik}q_{kl}q_{lj} + \cdots.
\end{equation}

Therefore, the matrix elements $f_{ij}$ of the fundamental matrix $F$ are the average number of times that a state $j$ appears in the chain if the process starts at state $i$:

\begin{equation}
f_{ij} = \sum_{n=0}^{\infty} q_{ij}^{(n)},
\end{equation}
where $q_{ij}^{(n)}$ is the $i,j$ element of the $Q^n$ matrix.

For all $i,j$, we can construct the fundamental matrix $F$ using the infinite sum formula:

\begin{equation}
F = \sum_{n=0}^{\infty} Q^n = (I - Q)^{-1},
\end{equation}
since $q_{ij}<1$ $\forall$ $i,j$, the expression above is valid.

Once the fundamental matrix is known, we can determine the number of times $n_{jk}$ that a transition $j\rightarrow k$ occurs if the cycle is started at state $i$, as the product between the number of times that state $j$ appears and the probability of moving from $j$ to $k$: $n_{jk} = f_{ij}p_{jk}$ (no Einstein summation rule). To find the average run length of kinesin starting the cycle at state $i$, $\langle x\rangle^{(i)}$, we define a step matrix $S$ whose elements are: $s_{jk}=0$ if the transition $j\rightarrow k$ does not result in a step; $s_{jk}=1$ if the transition $j \rightarrow k$ results in a step forward; and $s_{jk}=-1$ if a step backward occurs every time kinesin's state changes from $j$ to $k$. The average run length is:

\begin{equation}
\langle x\rangle^{(i)} = \sum_{j,k}n_{jk}s_{jk}\Delta x,
\end{equation}
where $\Delta x$ is the $8.2$ nm step size.

We know from Eq. 2 that the average time $\tau_j$ to leave a state $j$ is equal to $a_j^{-1} = 1/\sum_{j\ne k} \omega_{j\rightarrow k}$. The average run time $\langle t\rangle^{(i)}$ is therefore the sum over all states $j$ of the product of the total time to leave state $j$ and the number of times that state $j$ appears on average, if the Markov process starts at $i$:

\begin{equation}
\langle t\rangle^{(i)} = \sum_j f_{ij}\tau_j = \sum_j\frac{f_{ij}}{\sum_{k=1}^N \omega_{j\rightarrow k}}.
\end{equation}

%With the values of $\langle x\rangle^{(i)}$ and $\langle t\rangle^{(i)}$, we can calculate $v_{MC}$, as previously defined, in every step of the simulated annealing scheme, which is explained in detail in the next section.

Thus the run length and run time are calculated by two independent methods: kinetic Monte Carlo and Markov theory. %which we explain in detail in the following subsections. 
Markov theory is used within the optimization to calculate the observables numerically. Once the optimization step is finished, we use kinetic Monte Carlo to find error bars associated with the observables.

In Markov theory, we calculate the mean velocity $\langle v_{MC} \rangle= \langle x\rangle/\langle t\rangle$, from the mean run length $\langle x\rangle$, and mean run time $\langle t\rangle$. This velocity is the quantity calculated by \citep{Yajima2002}. It is important to point out that this speed is not the same as $v = \langle x/t \rangle$, that can be determined from kinetic Monte Carlo. However, in all our calculations the values of $v_{MC}$ and $v$ are within the experimental error bars in \citep{Schief2004}, which justifies the use of Markov theory within the optimization scheme.

\subsection*{Learning and Optimization Algorithm}

Within the 25-state kinetic model, the rate constants found elsewhere in the literature \citep{Herskowitz2010} are systematically changed by an optimization process until we obtain an agreement between our numerical results and the experimental data shown in \citep{Yajima2002,Schief2004}.

In addition to the states that are part of the kinetic model, the rate constants related to the particular mechanical and chemical transitions connecting these states play a vital role in the determination of the observables of the system, as explained above. Consequently, changes in the rate constants will be followed by changes in the speed and processivity of kinesin as a function of ATP and hydrolysis product concentrations. The rate constants provided in the literature via modeling for the 25-state model (c.f. Table 1, column ``Initial'' \citep{Herskowitz2010}) cannot reproduce correctly the experimental data on kinesin speed versus ATP concentrations at different experimental conditions. %Mention simulated annealing here
The present scheme systematically varies the rate constants in order to train the 25-state model to reproduce a subset of existing experimental results. The model is then tested against ``out-of-sample'' data to assess its robustness when presented with new data. %Add extra info on simulated annealing and machine learning.

We define our \emph{objective function} $\Omega$ as the Euclidean distance between the experimental and predicted measurements:

\begin{equation}
\Omega = \int_{y_1}^{y_2} dy |f_{exp}(y)-f_{num}(y)| ,
\end{equation}
where $f_{exp}$($y$) is the experimental data and  $f_{num}$($y$) is from the kinetic model in a given simulation. We further define a normalized objective function $\Omega^n$:

\begin{equation}
\Omega^n = \frac{1}{2N}\sum_{i=1}^{N}(|f_{exp}^n(y_{i+1})-f_{num}^n(y_{i+1})+f_{exp}^n(y_{i})-f_{num}^n(y_i)|),
\end{equation}
where $N$ is the number of data points, and:

\begin{eqnarray}
f_{exp}^n(y) & \equiv & \frac{f_{exp}(y)}{\max(f_{exp}(y),f_{num}(y))},\\
f_{num}^n(y) & \equiv & \frac{f_{num}(y)}{\max(f_{exp}(y),f_{num}(y))}.
\end{eqnarray}

Here the variable $y$ corresponds to an independent variable used in the experiment, such as ATP, ADP, and/or P$_i$ concentrations. The functions $f_{exp}(y)$ and $f_{num}(y)$ and the expression for $\Omega^n$ apply to any observable of the system, e.g. speed $v$ or processivity $x$. In practice, it is convenient to define a combined objective function $\phi^n(\{\Omega_{i,j}^n\})$, which simultaneously optimizes with respect to multiple observables $i$ (e.g. speed and processivity) under $j$ experimental conditions (nucleotide concentration, etc). The normalized function $\phi^n(\{\Omega_{i,j}^n\})$ is defined as:

\begin{equation}
\phi^n(\{\Omega_{i,j}^n\}) \equiv \frac{1}{n_on_s}\sum_{i=1}^{n_o}\sum_{j=1}^{n_s}\Omega_{i,j}^n,
\end{equation}
with $n_s$ equal to the number of experimental conditions, and $n_o$ is the number of observables. By construction, at $t=0$ the values of the objective function $\phi^n(\{\Omega_{i,j}^n\})$ lie in the interval [0,1]. For simplicity, from now on we refer to the normalized objective function as simply $\phi$.

Once an initial value of $\phi$ is determined for the original set of rate constants, we can start an optimization routine to minimize $\phi$ as a function of varying rate constants. The optimization utilizes simulated annealing. The system is started at a high temperature $T_{high}$ with the same initial rate constants used to determine $\phi$. We define $\beta$ as the inverse of temperature, and assume the Boltzmann constant $k_B=1$. We run the simulation 500 times, and in each step $m$, $\beta_m$ is varied in constant steps $\Delta\beta$, which consists in variable temperature steps that become smaller as $T_m\rightarrow 0$. At each temperature step, a Metropolis search is repeated 1000 times. The Metropolis step consists of selecting one of the rate constants at random, changing this rate constant, again based on a random procedure, and using a probabilistic criterion based on the current temperature $T_m$ and objective function $\phi$ to determine whether the change should be accepted or not.

The specific algorithm is as follows:

\begin{enumerate}

\item Initialize rate constants as given in Table 1, column ``Initial''.

\item Choose three random numbers: an integer $r_1 \in$ [1,$n$], where $n$ is the number of rate constants, a real $r_2 \in$ [0,1] and an integer $r_3 \in$ \{1,2\}. The first number $r_1$ is the index of the rate constant we are modifying, $k_{r_1}$. The change is done by:

\begin{equation}
k'_{r_1} = k_{r_1} + 0.25k_{r_1}r_2(-1)^{r_3},
\end{equation}
so that the new rate $k'_{r_1} \in [0.75k_{r_1},1.25k_{r_1}]$.

\item All the observables are recalculated for the new set of rate constants $\{k'_i\}$ using the Markov chain method.

\item A new objective function $\phi_{new}$ is determined and the acceptance probability is calculated as follows:

\begin{eqnarray}
P = \left\{
\begin{array}{ccc}
1& , & \textrm{if } \phi_{new}<\phi \\
 \\
e^{-|\phi-\phi_{new}|/T_m}& , & \textrm{otherwise.}
\end{array}\right.
\end{eqnarray}

If the change is accepted, $k_{r_1}\rightarrow k_{r_1}'$ and $\phi\rightarrow\phi_{new}$; otherwise the original rate constants remain the same. Return to 2 until final Metropolis step is concluded.

\item We record the lowest value of $\phi$ given in the 1000 Metropolis steps and the rate constants that generated this value, and use them as a starting point at the next temperature $T_{m+1}<T_m$. In some cases, we may want to fine tune the fitting, which is done by restarting the simulated annealing at $T_{high}$ but with the values of $\{k_i\}$ that were previously optimized. This prevents the parameters from being trapped at a local minimum and may result in a even smaller value of $\phi$ at the end of the second optimization.

\item Once the optimization is finished and the value of $\phi$ is low (of the order of 10$^{-3}$ or lower), we use the new set of rate constants to determine the observables now through the kinetic Monte Carlo method. For each data point, we run KMC 30 times for different random seeds, and all results are averaged over the realizations.
\end{enumerate}

The optimization is done for 100 different initial conditions, and each one will reach a different solution in parameter space. The solution that gives the best fit is the one we choose to model our system. We discuss our results and observations in the next sessions.

%\subsection*{Subsection}

%\subsubsection*{Subsubsection}

%Further text subdivisions can be made with the \emph{subsection} and
%\emph{subsubsection} commands.

\section*{\small{RESULTS}}

\subsection*{Model Implementation}

In order to assess the ability of our modeling to refine the values of the constituents rate constants in an unbiased way, we utilize a train/test machine learning approach. In each numerically optimized experiment described below, we perform the Markov chain/simulated annealing fitting procedure by initializing the entire set of rate constants, performing 100 realizations, and extracting the final ``best fit" rate constants within our mechanochemical model that best reproduce a specified set of training experimental data, measured under various [ATP], [ADP], and [P$_i$] concentrations. Once the optimized rate constants have been determined, they are held fixed, and we compute model predictions for additional experimental data {\it not used in the fit} in order to assess, without bias, the quality of the model and thus the computed rate constants.

The experimental data used include velocity vs. [ATP] curves from gliding motility assays performed by Schief {\it et al.} \citep{Schief2004} and processivity data from Yajima {\it et al.} \citep{Yajima2002}.  It is important to point out that Ref.\citep{Schief2004} shows data for wild-type kinesin and Ref.\citep{Yajima2002} studies the RK430G kinesin mutant. Kinesin RK430G presents the same speed as wild-type kinesin but its processivity is shorter \citep{Yajima2002}. For our model, the only rate affected for shorter processivity at the same speed is the detachment rate, ($k_{32}$ in Table 1). Therefore we can safely assume that all the other rate constants should be the same for both kinesin types. Once we have the optimized set of rate constants (shown in Table 1, column ``Optimized''), we run a KMC simulation to obtain the position of the molecule as a function of time at different ATP concentration. The $x(t)$ curves are fitted into a first-order polynomial in $t$ by a least-squares fit, for each independent KMC run (function {\it polyfit} in MATLAB). The speed of kinesin for that particular run is simply the slope of the fit. Error bars are obtained from KMC calculations, calculating the standard error of the mean over 30 realizations. It is important to emphasize that our objective function only comprises experimental data on processivity and speed, not constraining the number of ATP hydrolyzed \emph{per} step. During the kinetic Monte Carlo simulation, after the optimization, we can easily count how many ATP molecules are hydrolyzed every time kinesin steps forward.

\subsection*{Numerical Experiments}

Here we explain each experiment done throughout the optimization procedure, starting from curves obtained using the initial assumed rates. Table 2 shows the concentrations of ADP and P$_i$ in each numerical experiment performed (training phase).\\

\noindent{\bf 1. No optimization:} In this experiment, we use the rate constants compiled from the literature and calculated using the Wormlike chain model for the neck linker %[add new ref for this]
\citep{Herskowitz2010}. The rates are listed in Table 1, column ``Initial".  We try to reproduce the curves of speed for two cases, both at [P$_i$] = 0.0 mM and: [ADP] = 0.0 mM (Fig. 5A), and [ADP] = 1.0 mM (Fig. 5B). As observed, the initial rates fail to reproduce the expected curves for kinesin speed.\\

\noindent{\bf2. Experiment A:} In experiment A we include two experimental results in our training phase: [ADP] = 0.0 mM and [P$_i$] = 0.0 mM, and [ADP] = 1.0 mM and [P$_i$] = 0.0 mM, with speed data and processivity data from \citep{Schief2004, Yajima2002}. Once the optimization is completed, we obtain a set of optimized rate constants. We then use these new rates to restart the system at a higher temperature and redo the annealing procedure in order to find a lower value of the objective function. As there was no inorganic phosphate in solution during this training phase, the optimized rates cannot correctly predict the speed and processivity data for the experiments with inorganic phosphate in solution. We use this set of rates as the initial set to the following numerical experiment. \\

\noindent{\bf 3. Experiment B}: Now we change the training phase to reproduce experimental data from the following sets: [ADP] = 0.0 mM and [P$_i$] = 0.0 mM, and [ADP] = 1.0 mM and [P$_i$] = 5.0 mM. For the experimental set containing P$_i$ in solution, we assume that P$_i$ has no effect on the processivity \citep{Schief2004} and use the same processivity value for the experiment with [ADP] = 1.0 mM and [P$_i$] = 0.0 mM. The presence of inorganic phosphate is necessary to constrain the rates related to P$_i$ capture, and therefore, will in principle be able to reproduce experiments with nonzero [P$_i$] during the testing phase. The results of the training phase are shown in Fig. 6, green data (upper right corner).

\subsection*{Testing Phase}

After finding the optimized rates and how well they reproduce the data included in the objective function, one should test whether the rates are able to reproduce data that was not previously included in the optimization routine. The test phase is done for the data in \citep{Schief2004} for the three remaining sets: [ADP] = 5.0 mM, [P$_i$] = 0.0 mM; [ADP] = 0.0 mM, [P$_i$] = 10.0 mM; and [ADP] = 1.0 mM, [P$_i$] = 0.0 mM. The latter experimental condition was previously a training set, but since the final experiment did not constrain to it, it will now be part of the testing scheme. The kinetic Monte Carlo results of the test show an excellent agreement with the out-of-sample data for the experiments with no inorganic phosphate in solution. The experiment with [ADP] = 0.0 mM, [P$_i$] = 10.0 mM gives a high estimate of speed for low [ATP], but matches well experimental results for saturating ATP concentrations. The speed results associated to the testing phase are also shown in Fig. 6.

One important result of the optimization is that after Experiment B, the best results in both the training and testing phases recover the $\sim$ 1 ATP hydrolysis per step forward, as observed experimentally \citep{Coy1999,Yildiz2004}. Table 2 shows the ratio of ATP hydrolysis per step forward for each of the experiments and the testing phase after Experiment B.

%In addition, we also recover the ~1 ATP hydrolysis/step as confirmed by experimental evidence. The speed vs. ATP concentration for this experiment shown in Figure 6, and processivity and ATP hydrolysis results are shown in Table 4.

%The optimized rates result in an excellent fit with the out-of-sample data. In addition, we also recover the ~1 ATP hydrolysis/step as confirmed by experimental evidence.

\section*{\small{DISCUSSION}}

Mechanochemical cycles for kinesin have been widely proposed since the first experimental results that allowed a more detailed investigation of particular states of the kinesin walk. Our mechanochemical model, a general network of states, builds upon previously prescribed states and pathways for kinesin walk, based on both theory and experiments.

An early review by Vale and Millikan \citep{Vale2000} compares the kinetic cycles of kinesin and myosin by presenting a simplified mechanochemical model very similar to the cycle proposed by Rice {\it et al.} \citep{Rice1999}. The latter was the first work to identify the conformational changes in the neck linker as the force-generating action responsible for the mechanical step. A 1HB start state of empty-ADP was also present in this model, however it was not yet identified as the ATP-waiting state. As experiments extracted more information and gave stronger evidence for a hand-over-hand model of kinesin walk, proposed models became more complex,  adding more intermediate states between steps \citep{Schief2001, Carter2005} and possible branches \citep{Schief2001}. Experimental evidence also settled on the existence of certain states of the model for conventional kinesin, such as the ATP-waiting state \citep{Alonso2007} as the 1HB empty-ADP state, and the ADP-ADP detachment state \citep{Yajima2002}. On the other hand, some states were proven to be forbidden in conventional kinesin due to neck linker strain between the two heads \citep{Guydosh2009}. Some theoretical constructs were based largely on kinetic cycles assuming thermodynamic equilibrium fitted to experimental data. Such models were advanced by \citep{Hyeon2011,Liepelt2010,Czovek2011}. The works of Liepelt {\it et al}. \citep{Liepelt2007a,Liepelt2009,Liepelt2010} pioneered in presenting a clear distinction between which transitions are considered chemical or mechanical, yet treating them at the same theoretical level when building the mechanochemical model. They also introduced a network theory that offers multiple possible pathways for kinesin stepping. The presence of these shortcuts allows for backstepping and futile hydrolysis, such as found by Clancy {\it et al.} \citep{Clancy2011}, in which they propose an universal model with branches for a kinesin mutant.  Most of the kinetic cycles discussed above are subsets of our proposed network model, as shown in Fig. \ref{fig:result_fig10} for the pathways of Refs.~\citep{Schief2001,Liepelt2010,Hyeon2011,Clancy2011} (some of these cycles assume states that we have excluded from the model based on previous works \citep{Guydosh2009}). Indeed, further analysis of our results shows that the 25-state mechanochemical model proposed by our work is greatly reduced to a few most-visited states. However there are sporadic visits to rare states, which may become important as they are more or less visited with varying kinesin type and experimental conditions. Thus our model provides a systematic way to analyze the differences in kinesin walk under diverse conditions. 

Our results are analyzed once the set of optimized rates is determined via the machine learning algorithm presented in the Results session. The data analysis is done by recalculating the observables (speed and processivity) via kinetic Monte Carlo for 30 different random realizations.  Error bars are estimated from the standard error of the mean. A small number of samples is chosen in order to get error bars that are of the same order of magnitude as the ones in Schief {\it et al.} \citep{Schief2004}. Other information that is available through kinetic Monte Carlo are most visited states, time spent in each state and pathways. 

States that are frequently visited in our model mostly coincide with states proposed in other kinetic models  \citep{Cross2004,Moyer1998,Valentine2007,Schief2001,Hyeon2011}. In particular, we find that kinesin not always chooses a specific pathway, but instead takes alternative cycles at each step. Our dominant pathways are shown in Fig. 8. In the absence of hydrolysis products, we found no difference in pathways chosen for different ATP concentrations. In this scenario, about 55\% of the time that kinesin takes a step, it does so through pathway 23-11-1-3-5-19-23 in Fig. 8. In 40\% of the time, the pathway chosen is 23-11-1-2-4-6-23.
One of the main results from Ref.~\citep{Guydosh2009} is that kinesin spends about 93\% of the time with one head bound to the microtubule, in the ATP-waiting state, at low ATP concentration. We tested our model to check if we observe a similar trend of time spent at the ATP-waiting state. Our results are reported in Fig. 9. Indeed, at low ATP concentration, our results indicate that kinesin spends nearly 90\% of the time of the processive run at the ATP-waiting state (93\% at 2 $\mu$M [ATP]). As expected, as ATP concentration is increased, the time spent at the waiting state shortens. In terms of the actual time spent at the waiting state, our value is about 4 times shorter ($\approx$100ms) than the value found by \citep{Guydosh2009}. This is explained by our value of the ATP binding rate being almost 4 times higher than the one reported in \citep{Guydosh2009}. Both rates are within the accepted rates of 1-6 $\mu$M$^{-1}$ s$^{-1}$ \citep{Ma1995,Gilbert1994,Gilbert1995,Farrell2002,Auerbach2005}, and the discrepancy can be explained by the experimental conditions and systems (single molecule vs. gliding motility assay) studied. In fact, most of our rates are well within accepted values, even though they are allowed to vary without constraints. On the other hand, this also results in a few of our rates falling outside the range of experimental values. One advantage of this method is that we are able to determine rates which are not able to be found experimentally, especially those associated to very fast events such as kinesin binding/unbinding to the microtubule.

The change of speed and processivity of kinesin with added ADP is understood by analyzing the most common pathway that the protein undergoes during procession. Without the addition of extra ADP to the experiment, the waiting state of kinesin can only bind ATP, and the cycle of hydrolysis-stepping takes place continuously. As more ADP is added to the assay, the waiting state can bind either ADP or ATP. Whenever it binds ADP, it goes into an ADP-ADP state with one head bound to the microtubule, which is the state that leads to detachment. As this state is more visited, over time the chances of detaching becomes higher thus reducing processivity. Similarly, every time ADP binds to the waiting state instead of ATP, the protein must wait for ADP to unbind before ATP is free to bind and hydrolyse, thus slowing down the walk and reducing the speed.

The simulated annealing method has proven to fit well experimental data by finding new rate constants for the model chosen. The fact that our model is able to recover the rate of 1 ATP per step forward without constraining to it is a major strength of the procedure. We are also able to run very fast simulations by choosing to use Markov chain calculations within the optimization scheme, which are much simpler to solve in each Metropolis step as opposed to running kinetic Monte Carlo (Markov chain requires only a matrix inversion whilst KMC would require hundreds of independent realizations to calculate the average values of the observables). Furthermore, our finite Markov model does not require solving of Master equations to compute observables. A steady-state solution is also not required as kinesin is not allowed to rebind to the microtubule after it reaches the detached state, thus the verification of detailed balance is only necessary if we assume that the system operates near equilibrium. As we chose to solve our model with a minimum of assumptions, we do not constrain our rates to obey equilibrium conditions and therefore most of our transitions are practically irreversible, with reverse rates to some reactions close to zero, characterizing a non-equlibrium system. A difficulty that arises from running stochastic simulations for optimizing rate constants is that is it not always easy to choose which set of rates takes us to the best local minimum and still represent a {\it physical} system. Thus one must be careful to analyze the results by comparing the new rates to previously determined experimental rates, and constraining the simulation to more experimental data.  Moreover, to provide a more accurate quantitative description of rates, further investigation may be required, such as a sensitivity analysis of the optimized rate constants. Nonetheless, our method is successful in providing a qualitative picture of changes in pathways as experimental conditions change, as we have shown in this work, and enables systematic studies of different kinesin types.

\section*{\small{CONCLUSION}}

In summary, we have successfully implemented a novel simulated annealing method to optimize kinesin catalytic cycle rate constants by constraining kinetic Monte Carlo and Markov chain discrete state simulations to yield average processivity and velocities from experiment. The modeling yields a general network for the kinesin catalytic cycle, capable of making quantitative, verifiable predictions for independent experiments not included in the rate constant optimization procedure. The new discrete state model can be further extended to include additional states for modeling mutant kinesins, in order to elucidate the critical role of neck linker docking in kinesin mechanochemistry. The kinetic model also provides a potential link to the atomistic scale, by highlighting the discrete states of the kinetic cycle whose rate constants are maximally sensitive under optimization to environmental, structural, and chemical changes.
%The source file for this document is called
%\emph{bj\_latex\_template.tex}.  Apart from this \LaTeX\ file, you
%will also need the bibliography file, the \BibTeX\ style file, and the
%EPS and PDF figure files.

%See the bibliography file \emph{bj\_bibtex\_template.bib} for the
%literature data.  It was mostly generated from the saved
%text-formatted PubMed entries using the \emph{med2bib} program and
%edited by the \emph{tkbibtex} or directly in the \emph{emacs} editor.

%The \emph{biophysj.bst} file is a \BibTeX\ style file that contains
%information about the format required by Biophysical Journal for the
%list of references.

%Figure file \emph{fig\_1.eps} was generated by \emph{xfig} program and
%converted to PDF by the \emph{epstopdf} program.  Most data plotting
%is easily accomplished by the \emph{gnuplot} program.

% Compile and format the bibliography (bj_bibtex_template.bib BibTeX
% file must be present in the document directory)
%\bibliography{bj_bibtex_template}
\bibliography{kin3}

\begin{thebibliography}{42}
\providecommand{\url}[1]{\texttt{#1}}
\providecommand{\urlprefix}{ }

\bibitem[Coy et~al.(1999)Coy, Wagenbach, and Howard]{Coy1999}
Coy, D.~L., M.~Wagenbach, and J.~Howard, 1999.
\newblock Kinesin Takes One 8-nm Step for Each ATP That It Hydrolyzes.
\newblock \emph{J. Biol. Chem.} 274:3667--3671.

\bibitem[Yildiz et~al.(2004)Yildiz, Tomishige, Vale, and Selvin]{Yildiz2004}
Yildiz, A., M.~Tomishige, R.~D. Vale, and P.~R. Selvin, 2004.
\newblock Kinesin walks hand-over-hand.
\newblock \emph{Science} 303:676--678.

\bibitem[Hackney(1994)]{Hackney1994}
Hackney, D.~D., 1994.
\newblock Evidence for alternating head catalysis by kinesin during
  microtubule-stimulated ATP hydrolysis.
\newblock \emph{Proc. Natl. Acad. Sci. (USA)} 91:6865--9.

\bibitem[Asenjo et~al.(2006)Asenjo, Weinberg, and Sosa]{Asenjo2006}
Asenjo, A.~B., Y.~Weinberg, and H.~Sosa, 2006.
\newblock {Nucleotide binding and hydrolysis induces a disorder-order
  transition in the kinesin neck-linker region.}
\newblock \emph{Nat. Struct. Mol. Biol.} 13:648--54.

\bibitem[Brunnbauer et~al.(2012)Brunnbauer, Dombi, Ho, Schliwa, Rief, and
  \"{O}kten]{Brunnbauer2012}
Brunnbauer, M., R.~Dombi, T.-H. Ho, M.~Schliwa, M.~Rief, and Z.~\"{O}kten,
  2012.
\newblock {Torque generation of kinesin motors is governed by the stability of
  the neck domain.}
\newblock \emph{Mol. Cell} 46:147--58.

\bibitem[D\"{u}selder et~al.(2012)D\"{u}selder, Thiede, Schmidt, and
  Lak\"{a}mper]{Duselder2012}
D\"{u}selder, A., C.~Thiede, C.~F. Schmidt, and S.~Lak\"{a}mper, 2012.
\newblock {Neck-linker length dependence of processive Kinesin-5 motility.}
\newblock \emph{J. Mol. Biol.} 423:159--68.

\bibitem[Miyazono et~al.(2010)Miyazono, Hayashi, Karagiannis, Harada, and
  Tadakuma]{Miyazono2010}
Miyazono, Y., M.~Hayashi, P.~Karagiannis, Y.~Harada, and H.~Tadakuma, 2010.
\newblock {Strain through the neck linker ensures processive runs: a
  DNA-kinesin hybrid nanomachine study.}
\newblock \emph{EMBO J.} 29:93--106.

\bibitem[Rice et~al.(2003)Rice, Cui, Sindelar, Naber, Matuska, Vale, and
  Cooke]{Rice2003}
Rice, S., Y.~Cui, C.~Sindelar, N.~Naber, M.~Matuska, R.~Vale, and R.~Cooke,
  2003.
\newblock {Thermodynamic properties of the kinesin neck-region docking to the
  catalytic core.}
\newblock \emph{Biophys. J.} 84:1844--54.

\bibitem[Rice et~al.(1999)Rice, Lin, Safer, Hart, Naber, Carragher, Cain,
  Pechatnikova, Wilson-Kubalek, Whittaker, Pate, Cooke, Taylor, Milligan, and
  Vale]{Rice1999}
Rice, S., A.~W. Lin, D.~Safer, C.~L. Hart, N.~Naber, B.~O. Carragher, S.~M.
  Cain, E.~Pechatnikova, E.~M. Wilson-Kubalek, M.~Whittaker, E.~Pate, R.~Cooke,
  E.~W. Taylor, R.~A. Milligan, and R.~D. Vale, 1999.
\newblock {A structural change in the kinesin motor protein that drives
  motility.}
\newblock \emph{Nature} 402:778--84.

\bibitem[Carter and Cross(2005)]{Carter2005}
Carter, N.~J., and R.~A. Cross, 2005.
\newblock {Mechanics of the kinesin step.}
\newblock \emph{Nature} 435:308--12.

\bibitem[Yildiz et~al.(2008)Yildiz, Tomishige, Gennerich, and Vale]{Yildiz2008}
Yildiz, A., M.~Tomishige, A.~Gennerich, and R.~D. Vale, 2008.
\newblock {Intramolecular strain coordinates kinesin stepping behavior along
  microtubules.}
\newblock \emph{Cell} 134:1030--41.

\bibitem[Moyer et~al.(1998)Moyer, Gilbert, and Johnson]{Moyer1998}
Moyer, M.~L., S.~P. Gilbert, and K.~a. Johnson, 1998.
\newblock {Pathway of ATP hydrolysis by monomeric and dimeric kinesin.}
\newblock \emph{Biochemistry} 37:800--13.

\bibitem[Cross(2004)]{Cross2004}
Cross, R.~A., 2004.
\newblock {The kinetic mechanism of kinesin.}
\newblock \emph{Trends in Biochem. Sci.} 29:301--9.

\bibitem[Clancy et~al.(2011{\natexlab{a}})Clancy, Behnke-Parks, Andreasson,
  Rosenfeld, and Block]{Clancy2011a}
Clancy, B.~E., W.~M. Behnke-Parks, J.~O.~L. Andreasson, S.~S. Rosenfeld, and
  S.~M. Block, 2011.
\newblock {A universal pathway for kinesin stepping.}
\newblock \emph{Nat. struct. mol. biol.} 18:1020--7.

\bibitem[Hyeon and Onuchic(2011)]{Hyeon2011}
Hyeon, C., and J.~N. Onuchic, 2011.
\newblock {A structural perspective on the dynamics of kinesin motors.}
\newblock \emph{Biophys. J.} 101:2749--59.

\bibitem[Liepelt and Lipowsky(2007{\natexlab{a}})]{Liepelt2007}
Liepelt, S., and R.~Lipowsky, 2007.
\newblock {Kinesin's network of chemomechanical motor cycles}.
\newblock \emph{Phys. Rev. Lett.} 98:258102.

\bibitem[Liepelt and Lipowsky(2009)]{Liepelt2009}
Liepelt, S., and R.~Lipowsky, 2009.
\newblock {Operation modes of the molecular motor kinesin}.
\newblock \emph{Phys. Rev. E} 79:011917.

\bibitem[Liepelt and Lipowsky(2010)]{Liepelt2010}
Liepelt, S., and R.~Lipowsky, 2010.
\newblock {Impact of slip cycles on the operation modes and efficiency of
  molecular motors}.
\newblock \emph{J. Stat. Phys.} 141:1--16.

\bibitem[Goulet et~al.(2014)Goulet, Major, Jun, Gross, Rosenfeld, and
  Moores]{Goulet2014}
Goulet, A., J.~Major, Y.~Jun, S.~P. Gross, S.~S. Rosenfeld, and C.~A. Moores,
  2014.
\newblock Comprehensive structural model of the mechanochemical cycle of a
  mitotic motor highlights molecular adaptations in the kinesin family.
\newblock \emph{Proc. Natl. Acad. Sci. USA.} 111:1837--1842.

\bibitem[Yajima et~al.(2002)Yajima, Alonso, Cross, and Toyoshima]{Yajima2002}
Yajima, J., M.~C. Alonso, R.~A. Cross, and Y.~Y. Toyoshima, 2002.
\newblock {Direct long-term observation of kinesin processivity at low load.}
\newblock \emph{Curr. Biol.} 12:301--6.

\bibitem[Sindelar and Downing(2007)]{Sindelar2007}
Sindelar, C.~V., and K.~H. Downing, 2007.
\newblock The beginning of kinesin's force-generating cycle visualized at
  9-\AA\ resolution.
\newblock \emph{J. Cell. Biol.} 177:377--385.

\bibitem[Kikkawa and Hirokawa(2006)]{Kikkawa2006}
Kikkawa, M., and N.~Hirokawa, 2006.
\newblock High-resolution cryo-EM maps show the nucleotide binding pocket of
  KIF1A in open and closed conformations.
\newblock \emph{EMBO J.} 25:4187--4194.

\bibitem[Guydosh and Block(2009)]{Guydosh2009}
Guydosh, N., and S.~Block, 2009.
\newblock Direct observation of the binding state of the kinesin head to the
  microtubule.
\newblock \emph{Nature} 461:125--128.

\bibitem[Schief and Howard(2001)]{Schief2001}
Schief, W.~R., and J.~Howard, 2001.
\newblock Conformational changes during kinesin motility.
\newblock \emph{Curr. Op. Cell Biol.} 13:19--28.

\bibitem[Clancy et~al.(2011{\natexlab{b}})Clancy, Behnke-Parks, Andreasson,
  Rosenfeld, and Block]{Clancy2011}
Clancy, B.~E., W.~M. Behnke-Parks, J.~O.~L. Andreasson, S.~S. Rosenfeld, and
  S.~M. Block, 2011.
\newblock {A universal pathway for kinesin stepping. Supplementary Materials}.
\newblock \emph{Nature structural \& molecular biology} 18:1020--7.

\bibitem[Herskowitz and Koch(2010)]{Herskowitz2010}
Herskowitz, L.~J., and S.~J. Koch, 2010.
\newblock {Discrete state model for Kinesin-1 with rate constants modulated by
  neck linker Tension}.
\newblock \emph{Nature Preced.} hdl:10101/npre.2010.5038.1.

\bibitem[Schwartz(2008)]{Schwartz2008}
Schwartz, R., 2008.
\newblock Biological Modeling and Simulation.
\newblock The MIT Press, Cambridge, Mass., 1 edition.

\bibitem[Gillespie(2007)]{Gillespie2007}
Gillespie, D.~T., 2007.
\newblock {Stochastic simulation of chemical kinetics.}
\newblock \emph{Annu. Rev. Phys. Chem.} 58:35--55.

\bibitem[Sumathy and Satyanarayana(2012)]{Sumathy2012}
Sumathy, S., and S.~V.~M. Satyanarayana, 2012.
\newblock {Dynamics of kinesin: A computational study}.
\newblock \emph{AIP Conf. Proc.} 1447:203--204.

\bibitem[Kutys et~al.(2010)Kutys, Fricks, and Hancock]{Kutys2010}
Kutys, M.~L., J.~Fricks, and W.~O. Hancock, 2010.
\newblock {Monte Carlo analysis of neck linker extension in kinesin molecular
  motors.}
\newblock \emph{PLoS Comp. Biol.} 6:e1000980.

\bibitem[Hughes et~al.(2011)Hughes, Hancock, and Fricks]{Hughes2011}
Hughes, J., W.~O. Hancock, and J.~Fricks, 2011.
\newblock {A matrix computational approach to kinesin neck linker extension.}
\newblock \emph{J. Theor. Biol.} 269:181--94.

\bibitem[Schief et~al.(2004)Schief, Clark, Crevenna, Howard, and
  Spudich]{Schief2004}
Schief, W.~R., R.~H. Clark, A.~H. Crevenna, J.~Howard, and J.~A. Spudich, 2004.
\newblock {Inhibition of kinesin motility by ADP and phosphate supports a
  hand-over-hand mechanism}.
\newblock \emph{Proc. Natl. Acad. Sci. USA.} 101:1183--1188.

\bibitem[Vale and MIllikan(2000)]{Vale2000}
Vale, R.~D., and R.~A. MIllikan, 2000.
\newblock The Way Things Move: Looking Under the Hood of Molecular Motor
  Proteins.
\newblock \emph{Science} 288:88--95.

\bibitem[Alonso et~al.(2007)Alonso, Drummond, Kain, Hoeng, Amos, and
  Cross]{Alonso2007}
Alonso, M.~C., D.~R. Drummond, S.~Kain, J.~Hoeng, L.~Amos, and R.~A. Cross,
  2007.
\newblock An ATP Gate Controls Tubulin Binding by the Tethered Head of
  Kinesin-1.
\newblock \emph{Science} 316:120--123.

\bibitem[Cz\"{o}vek et~al.(2011)Cz\"{o}vek, Sz\"{o}llosi, and
  Der\'{e}nyi]{Czovek2011}
Cz\"{o}vek, A., G.~J. Sz\"{o}llosi, and I.~Der\'{e}nyi, 2011.
\newblock {Neck-linker docking coordinates the kinetics of kinesin's heads.}
\newblock \emph{Biophysical journal} 100:1729--36.

\bibitem[Liepelt and Lipowsky(2007{\natexlab{b}})]{Liepelt2007a}
Liepelt, S., and R.~Lipowsky, 2007.
\newblock Steady-state balance conditions for molecular motors cycles and
  stochastic nonequilibrium processes.
\newblock \emph{EPL (Europhysics Letters)} 77:50002.

\bibitem[Valentine and Gilbert(2007)]{Valentine2007}
Valentine, M.~T., and S.~P. Gilbert, 2007.
\newblock To step or not to step? How biochemistry and mechanics influence
  processivity in Kinesin and Eg5.
\newblock \emph{Curr. Op. Cell Biol.} 19:75--81.

\bibitem[Ma and Taylor(1995)]{Ma1995}
Ma, Y.~Z., and E.~W. Taylor, 1995.
\newblock Kinetic Mechanism of Kinesin Motor Domain.
\newblock \emph{Biochemistry} 34:13233–13241.

\bibitem[Gilbert and Johnson(1994)]{Gilbert1994}
Gilbert, S.~P., and K.~a. Johnson, 1994.
\newblock {Pre-steady-state kinetics of the microtubule-kinesin ATPase.}
\newblock \emph{Biochemistry} 33:1951--60.

\bibitem[Gilbert et~al.(1995)Gilbert, Webb, Brune, and Johnson]{Gilbert1995}
Gilbert, S.~P., M.~R. Webb, M.~Brune, and K.~A. Johnson, 1995.
\newblock Pathway of processive ATP hydrolysis by kinesin.
\newblock \emph{Biochemistry} 373:671–676.

\bibitem[Farrell et~al.(2002)Farrell, Mackey, Klumpp, and Gilbert]{Farrell2002}
Farrell, C.~M., A.~T. Mackey, L.~M. Klumpp, and S.~P. Gilbert, 2002.
\newblock The Role of ATP Hydrolysis for Kinesin Processivity.
\newblock \emph{J. Biol. Chem.} 277:17079--17087.

\bibitem[Auerbach and Johnson(2005)]{Auerbach2005}
Auerbach, S.~D., and K.~A. Johnson, 2005.
\newblock Alternating Site ATPase Pathway of Rat Conventional Kinesin.
\newblock \emph{J. Biol. Chem.} 280:37048--37060.

\end{thebibliography}

\clearpage

\begin{table}[ht]
\caption{Rate constants used in Kinetic Model}
\vspace{0.3cm}
\centering
  \begin{tabular}{ c  l   l   l  l  }
    \hline \hline
Rate constants & $\{k_i\}$ (s$^{-1}$) & Initial${^1}$ &  Optimized${^2}$ & Transition    \\ \hline
 & $k_1$ &  500.0  & 384.38 & ADP unbinding \\
 & $k_2$ & 0.25 $\mu$M$^{-1}$ & 1.18$\times 10^{-8}$ $\mu$M$^{-1}$  & ADP binding   \\
 & $k_3$ &  800.0 & 630.13  & ATP hydrolysis     \\
Chemical & $k_4$ &  25.0 & 1.34 $\times 10^{-9}$  & ATP synthesis  \\
Rates & $k_5$ &  20.0 $\mu$M$^{-1}$ & 1.34 $\mu$M$^{-1}$  & P$_i$ absorption   \\
(2HB) & $k_6$ & 25.0 & 8359.33  & P$_i$ release  \\
 & $k_7$ & 5.0 & 1.75 $\times 10^{-4}$  & ADP unbinding   \\
 & $k_8$ & 0.3 $\mu$M$^{-1}$   & 1.47 $\times 10^{-10}$  $\mu$M$^{-1}$ & ATP binding   \\
 & $k_9$ & 50.0 & 6.61 $\times 10^{-8}$   & ATP unbinding  \\  \hline
 & $k_{10}$ & 10.0 & 1.12 $\times 10^{-3}$ & ATP hydrolysis    \\
 & $k_{11}$ & 25.0 & 2.34 $\times 10^{-2}$  & ATP synthesis   \\
 & $k_{12}$ & 250 & 5.20 $\times 10^{-5}$  & P$_i$ release   \\
&  $k_{13}$ & 20.0 $\mu$M$^{-1}$  & 8.62 $\times 10^{-4}$ $\mu$M$^{-1}$  & P$_i$ absorption  \\
& $k_{14}$ & 0.002 & 1.23 $\times 10^{-9}$   & ADP unbinding  \\
 & $k_{15}$ & 6.0 $\mu$M$^{-1}$  & 3.42 $\times 10^{-7}$ $\mu$M$^{-1}$ & ADP binding      \\
Chemical  & $k_{16}$ & 4.0 $\mu$M$^{-1}$  & 4.96 $\times 10^{-8}$ $\mu$M$^{-1}$    & ATP binding   \\
Rates  & $k_{17}$ & 50.0 & 0.41   & ATP unbinding  \\
(1HB) & $k_{18}$ & 25 & 1.00 $\times 10^{-4}$  & ATP synthesis  \\
 & $k_{19}$ & 400.0 & 6.00 $\times 10^{-3}$   & ATP hydrolysis  \\
 & $k_{20}$ & 20.0 $\mu$M$^{-1}$  & 2.34 $\times 10^{-4}$ $\mu$M$^{-1}$  & P$_i$ absorption    \\
 & $k_{21}$ & 81.0 & 3.48 $\times 10^{-4}$ & P$_i$ release    \\
 & $k_{22}$ & 1.5 $\mu$M$^{-1}$  & 3.82 $\mu$M$^{-1}$   & ADP binding    \\
 & $k_{23}$ & 110.0 & 149.44  & ADP unbinding   \\
 & $k_{24}$ & 4.0 $\mu$M$^{-1}$ & 4.65 $\mu$M$^{-1}$  & ATP binding   \\
 & $k_{25}$ & 80.0 & 2.30 $\times 10^{-2}$   & ATP unbinding  \\  \hline
Binding & $k_{26}$ & 4.90 $\times 10^{4}$ & 7.85 $\times 10^{5}$ &   NL docked-undocked  \\
 Mechanical Rates & $k_{27}$ & 36.75 & 1.12 $\times 10^{-6}$  &  NL undocked-undocked\\ \hline
 & $k_{28}$ & 0.7661 & 5.46 $\times 10^{-3}$   & APO, NL docked-undocked  \\
Unbinding & $k_{29}$ & 2736.0 & 2.36 $\times 10^{-3}$ & ADP, NL docked-undocked \\
Mechanical & $k_{30}$ & 246.0 & 1.12$\times 10^{-1}$  & APO, NL undocked-undocked  \\
Rates & $k_{31}$ & 8.78 $\times 10^{5}$ & 732.82   & ADP, NL undocked-undocked \\
 & $k_{32}$ & 250.0 & 3.71 & ADP, Detach from MT \\ \hline\hline

 \end{tabular}
\label{table:rates}
\small{${^1}$Ref. \citep{Herskowitz2010} cites all chemical rates (see references therein) and calculates all binding and unbinding mechanical rates.}
\small{${^2}$Rate constants obtained after Experiment B.}
\end{table}

%\clearpage

%\begin{table}[ht]
%\caption{ADP and P$_i$ concentrations for each numerical experiment performed}
%\vspace{0.3cm}
%\centering
 %\begin{tabular}{ r  c  c }
 %   \hline \hline
%Experiment & [ADP] (mM) & [P$_i$] (mM) \\\hline
%Initial Rate Constants & 0.0 & 0.0 \\
%Initial Rate Constants & 1.0 & 0.0 \\
%Experiment A & 0.0 & 0.0 \\
%Experiment A & 1.0 & 0.0 \\
%Experiment B & 0.0 & 0.0 \\
%Experiment B & 1.0 & 5.0 \\ \hline \hline
%\end{tabular}
%\label{table:exps}
%\end{table}

\clearpage

\begin{table}[ht]
\caption{Characteristics of initial, training, and testing numerical experiments. [ADP] and [P$_i$] are input; computed ATP hydrolysis reactions/step is determined as average ratio at 2.0 mM [ATP].}
\vspace{0.3cm}
\centering
 \begin{tabular}{ l  c   c   c }
   \hline \hline
Experiment & [ADP] (mM) & [P$_i$] (mM) & Computed ATP hydrolysis \\
& & &   reactions/step \\ \hline
Initial Rate Constants & 0.0 & 0.0 &  0.003\\
Initial Rate Constants & 1.0 & 0.0 & 0.003\\
Experiment A &  0.0 & 0.0 & 0.997 \\
Experiment A & 1.0 & 0.0 & 0.997\\
Experiment B & 0.0 & 0.0 & 0.999\\
Experiment B & 1.0 & 5.0 & 0.993\\
Testing phase & 0.0 & 10.0 & 0.995\\
Testing phase & 1.0 & 0.0 & 1.0 \\
Testing phase & 5.0 & 0.0 &  1.0 \\ \hline \hline
\end{tabular}
\label{table:rates}
\end{table}

\clearpage
\begin{figure}
   \begin{center}
      \includegraphics*[width=5.0in]{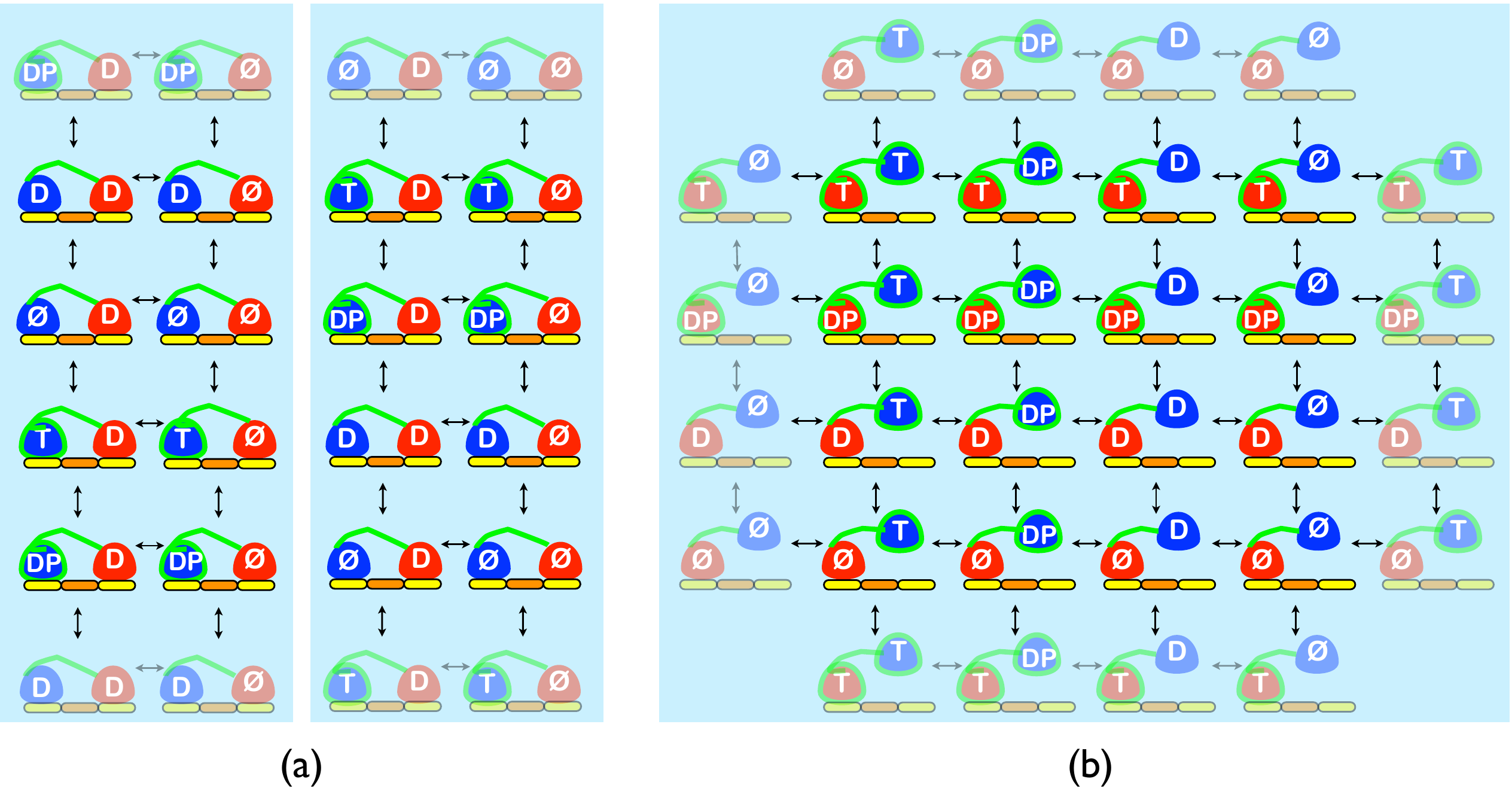}
      \caption{{ \bf Chemical transitions.} Alternating yellow-orange ovals correspond to $\alpha$- and $\beta$- tubulin strands comprising the microtubule. There are 24 allowed MT-bound states: 8 two-head bound states (a) and 16 single-head bound states (b).  Red designates front heads; and blue, trailing or unbound heads. Labels indicate the type of nucleotide bound to the head: ATP (T), ADP (D), ADP+P$_i$ (DP) or empty (APO state, \o). The neck linker (NL) is represented in green. If the neck linker is docked, the head is outlined in green. In the present model, all ATP and ADP+P$_i$ heads have a docked NL, whereas ADP and empty heads have an undocked NL. Shaded states illustrate periodic boundary conditions in the ``$x$'' and ``$y$'' directions of the state arrays. The 8 states on the right blue square are the same states as in the left square, but offset by two rows. The states are duplicated in order to facilitate visualization of the mechanical transitions in Fig. 3. (b) Chemical transitions between the 16 allowed one-head-bound states. These transitions also present periodic boundary conditions along both the ``$x$'' and ``$y$''  directions. }
      \label{fig:result_fig1}
   \end{center}
\end{figure}

\clearpage
\begin{figure}
   \begin{center}
\mbox{\subfigure{\includegraphics[width=3.25in]{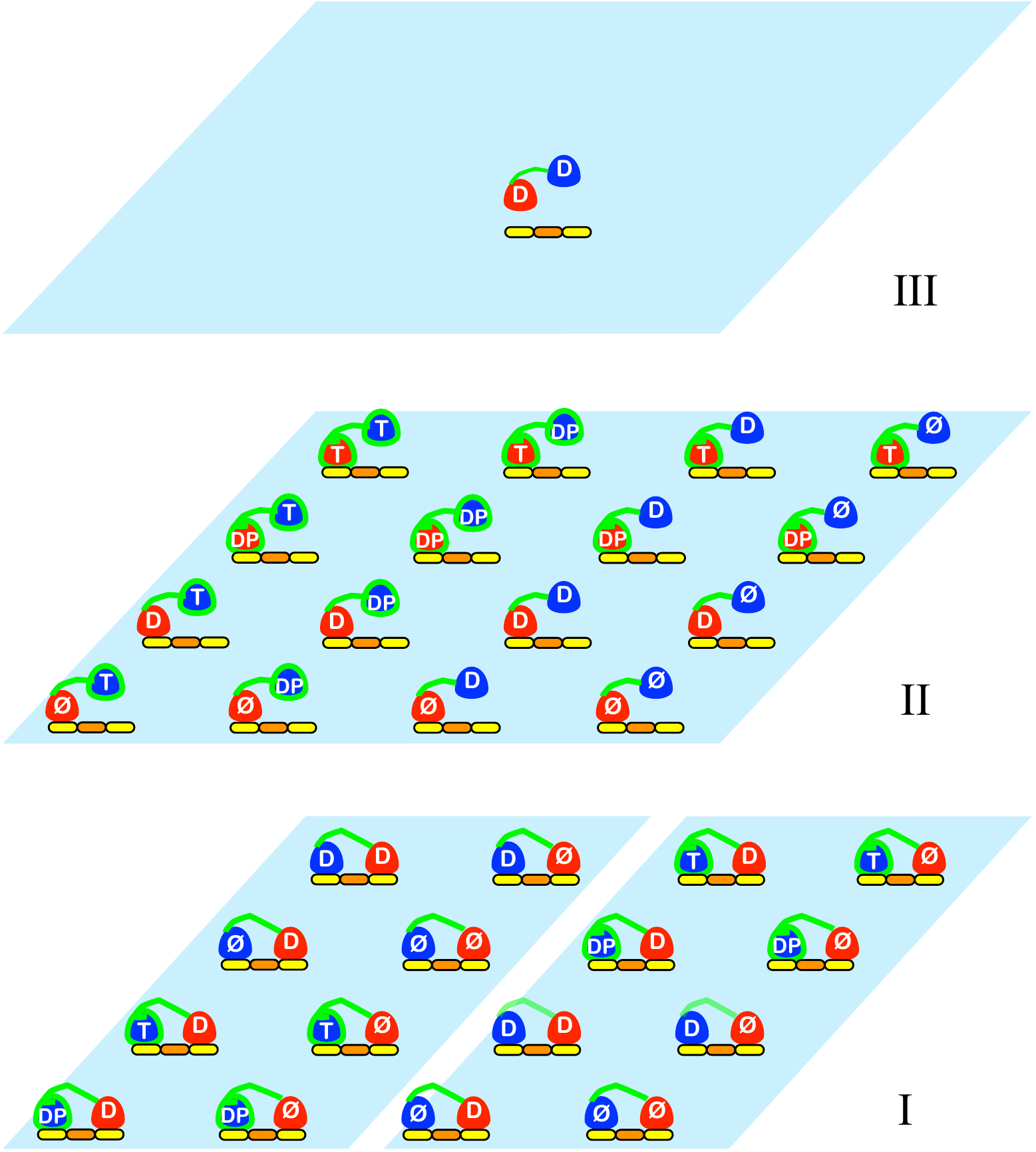}}\quad
\subfigure{\includegraphics[width=3.25in]{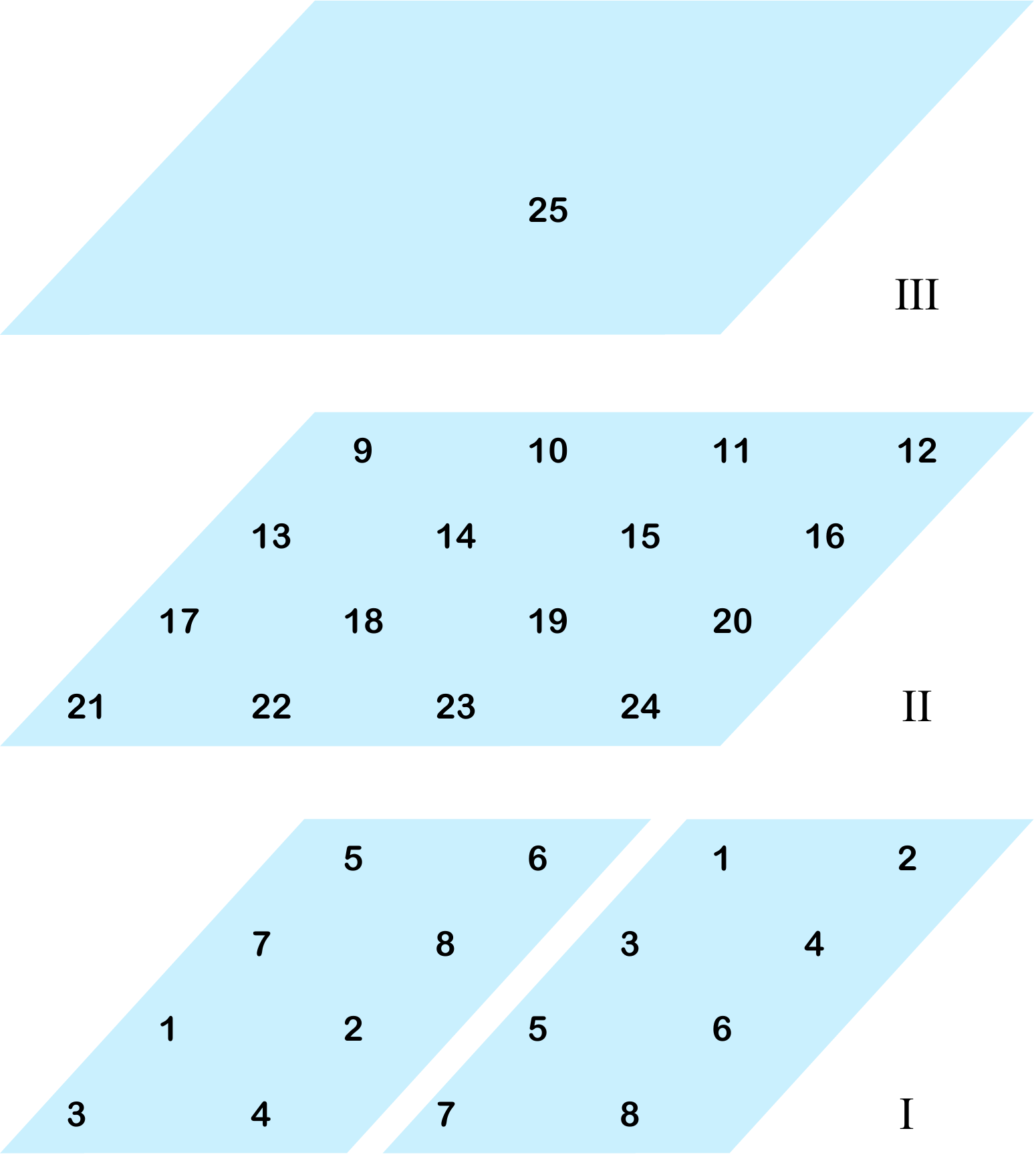} }}
      \caption{{\bf 25-state Kinetic Model.} Left: The MT-bound states from Fig. 1 (I, II) with the addition of a third level (III) consisting of a single, totally detached ADP-ADP state. These 25 states comprise all allowed transition states. Right: State labels corresponding to the kinetic model represented at left.}
      \label{fig:result_fig2}
   \end{center}
\end{figure}

%\clearpage
%\begin{figure}
%   \begin{center}
%      \includegraphics*[width=4.5in]{layers0}
%      \caption{Label of states in Figure 2. }
%      \label{fig:result_fig3}
%   \end{center}
%\end{figure}

\clearpage
\begin{figure}
\centering
\vspace{-0.5cm}
\mbox{\subfigure{\includegraphics[width=2.8in]{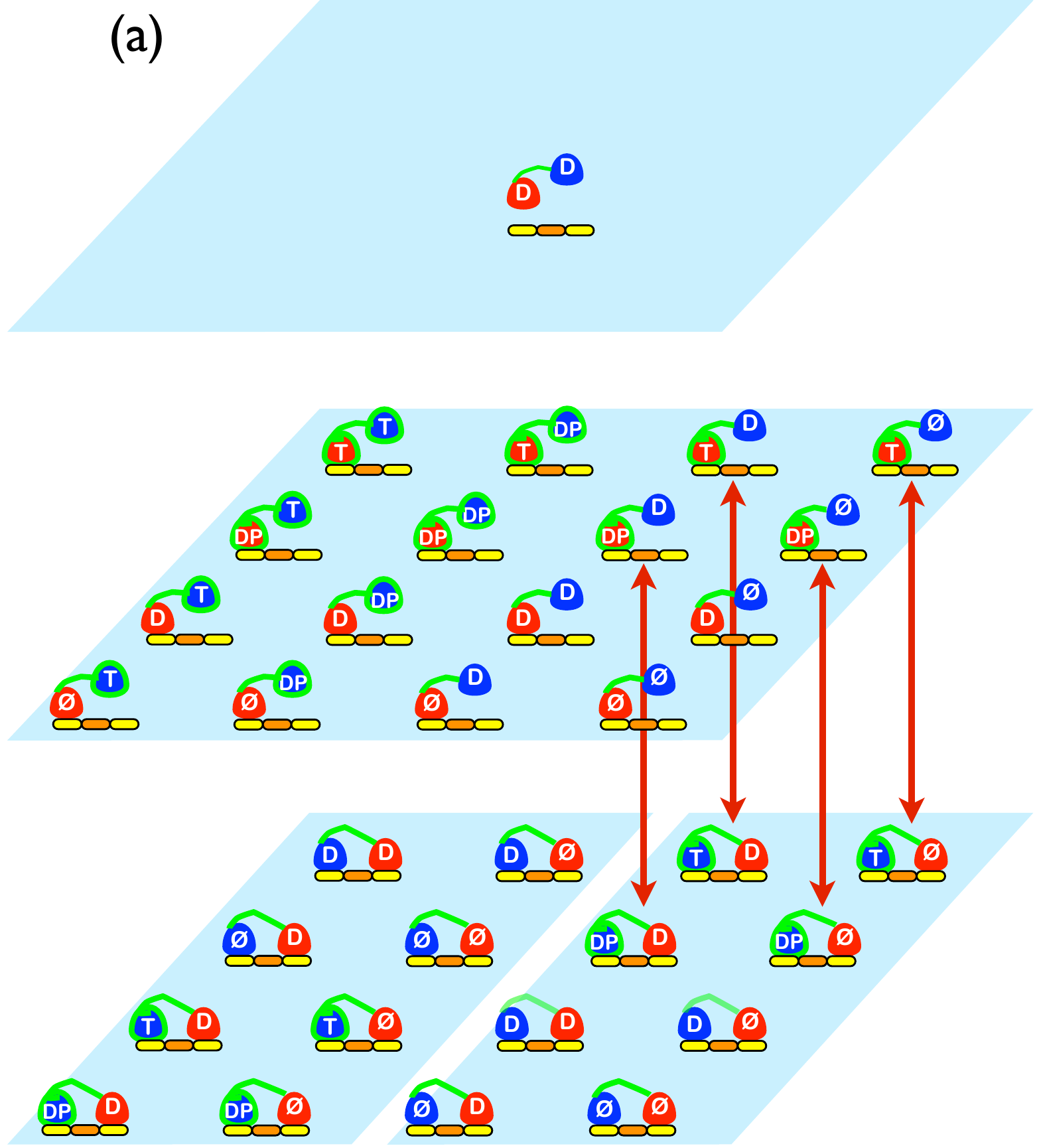}}\quad
\subfigure{\includegraphics[width=2.8in]{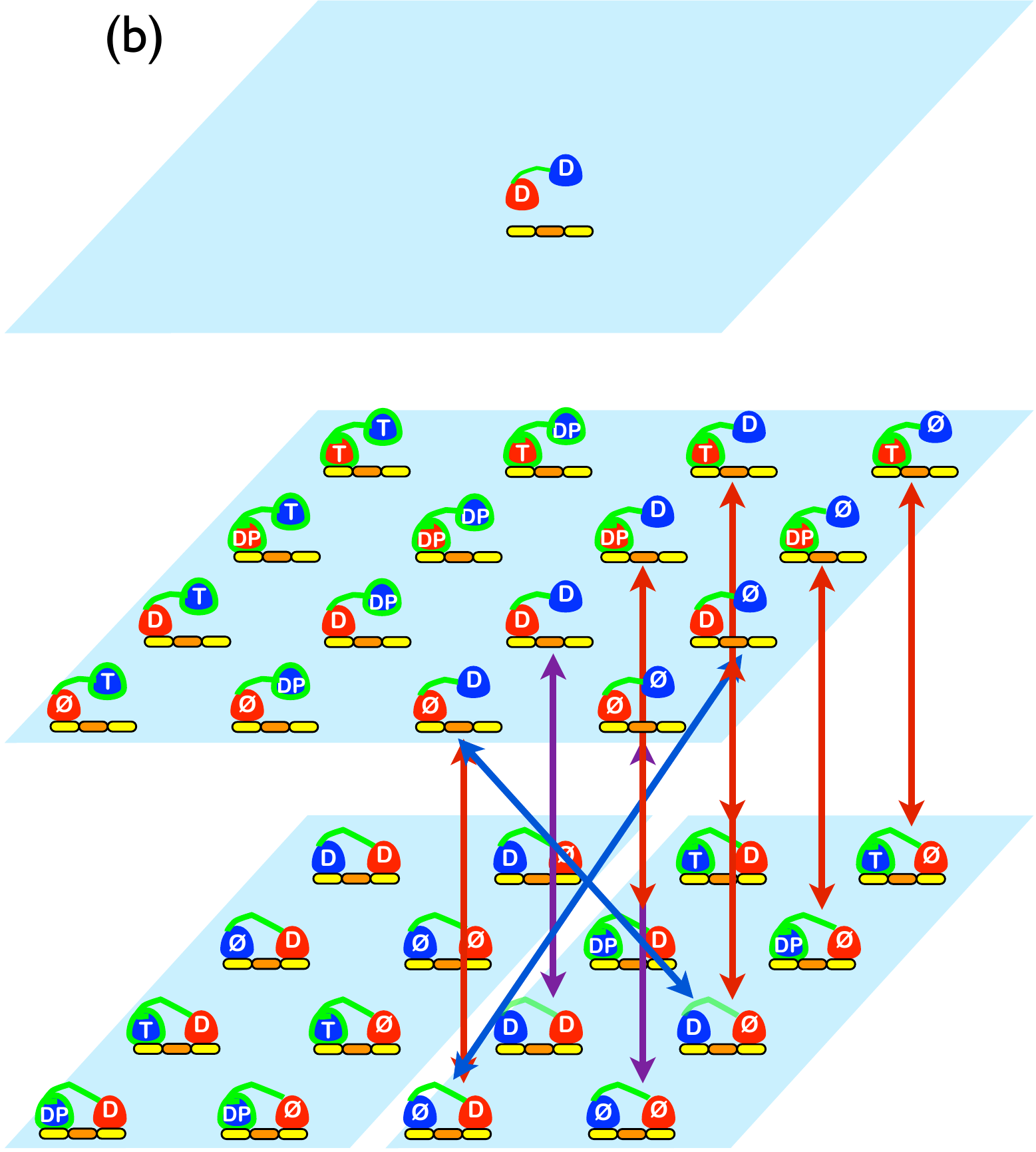} \vspace{2.5cm}}}
\end{figure}
\begin{figure}
\centering
\vspace{0.0cm}
\mbox{\subfigure{\includegraphics[width=2.8in]{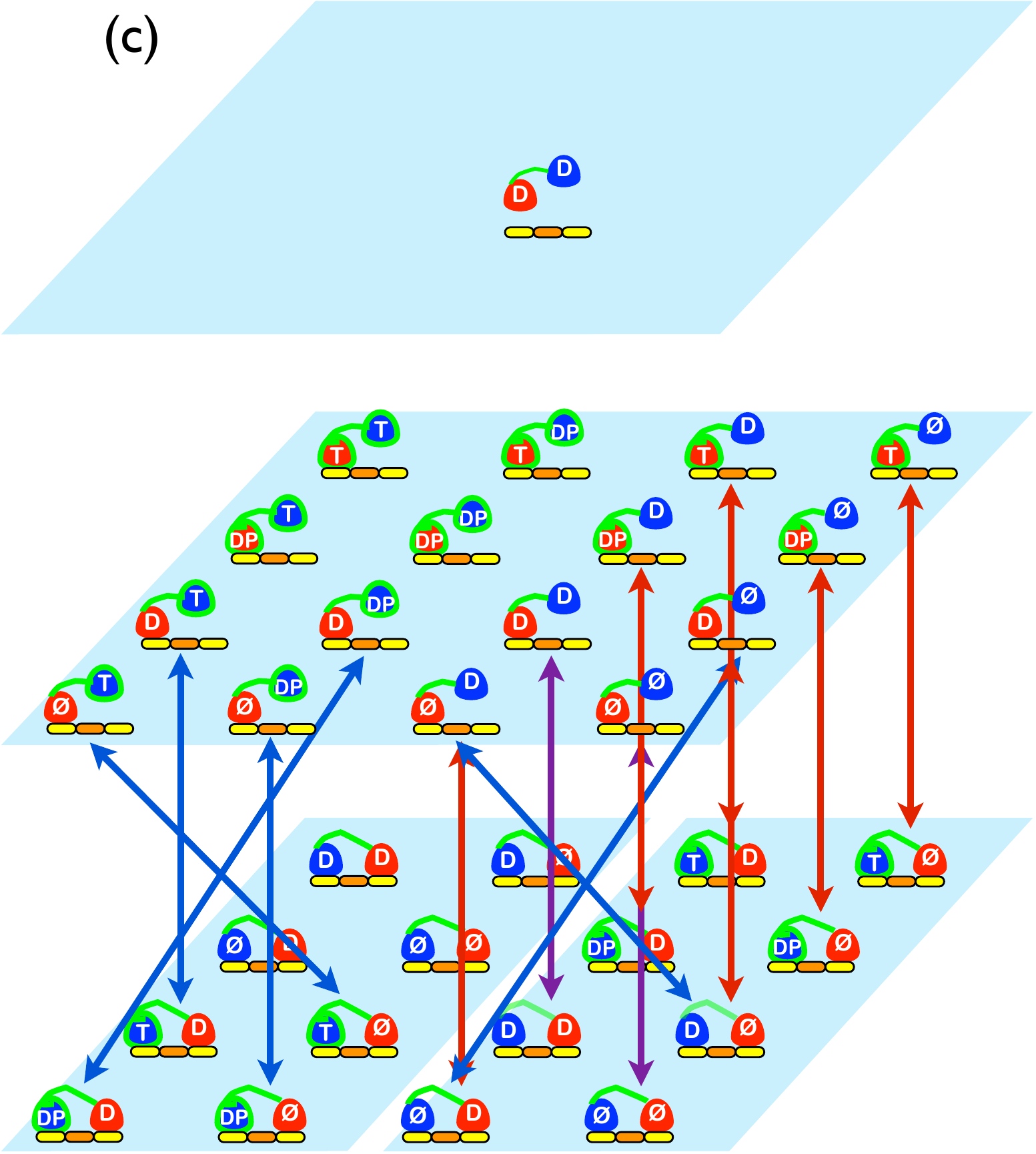}}\quad
\subfigure{\includegraphics[width=2.8in]{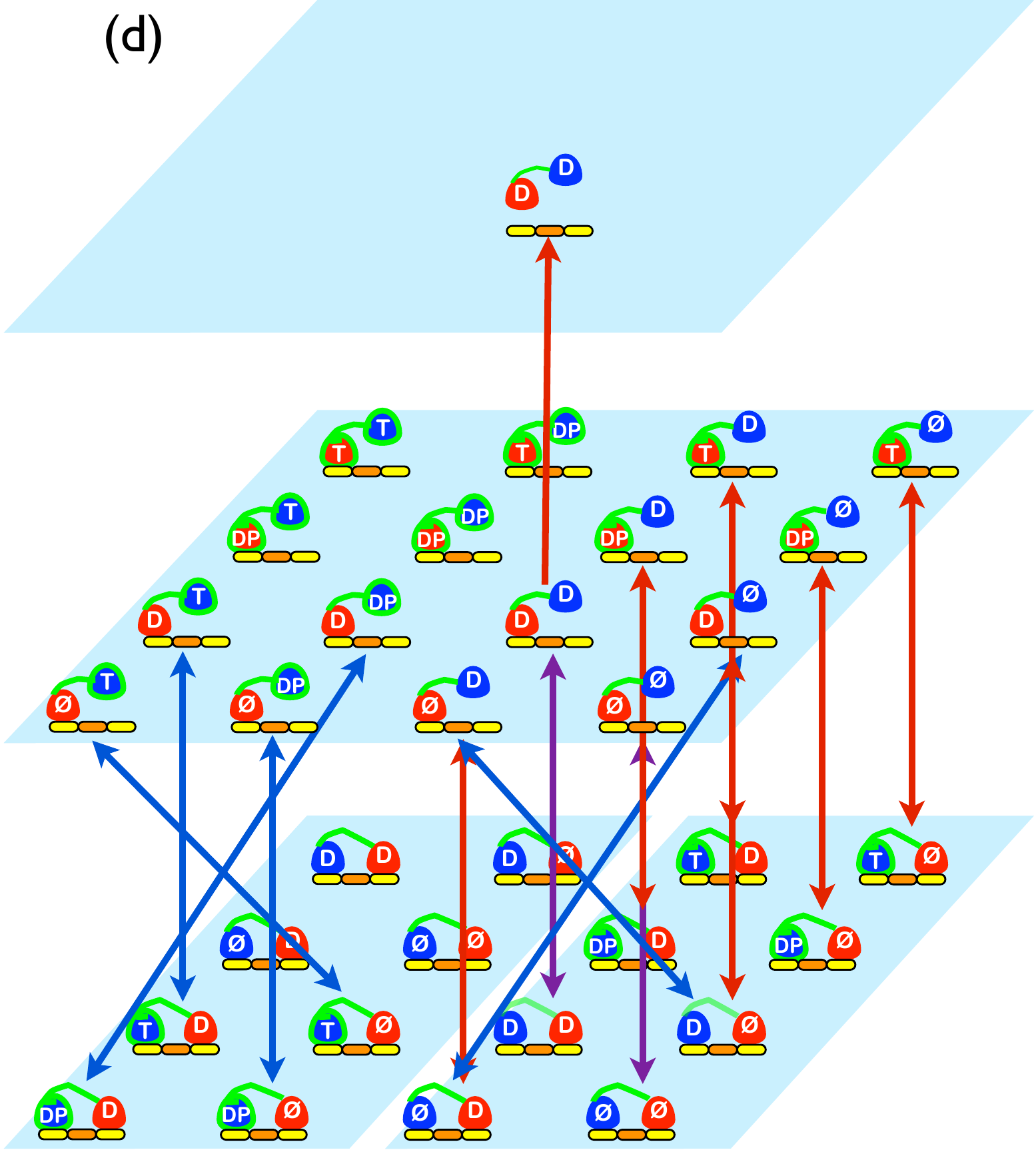} }}
\caption{{\bf Mechanical transitions between levels I, II and III of Fig. 2}.  (a)-(c): Red bi-directional arrows indicate unbinding of front head (red) (I$\rightarrow$II, $\Delta x = -8.2$ nm) or attachment of unbound head (blue) (II$\rightarrow$I, $\Delta x = 8.2$ nm). Blue bi-directional arrows indicate unbinding of trailing (blue) head (I$\rightarrow$II, $\Delta x = 0$ nm) or attachment of unbound head (blue) backward (II$\rightarrow$I, $\Delta x = 0$ nm). Purple bi-directional arrows: In level I, both heads have the same probability of unbinding. Unbinding of the front head implies a $\Delta x = -8.2$ nm step, and unbinding of the trailing head does not change total displacement. In level II, the unbound head has an equal probability of binding forward or backward, taking a $\Delta x = 8.2$ nm step when binding forward, and $\Delta x = 0$ nm when binding backward. (d): Unique detachment transition from the one-head-bound ADP-ADP state.  } \label{fig12}
\end{figure}

\clearpage
\begin{figure}
   \begin{center}
      \includegraphics*[width=6.5in]{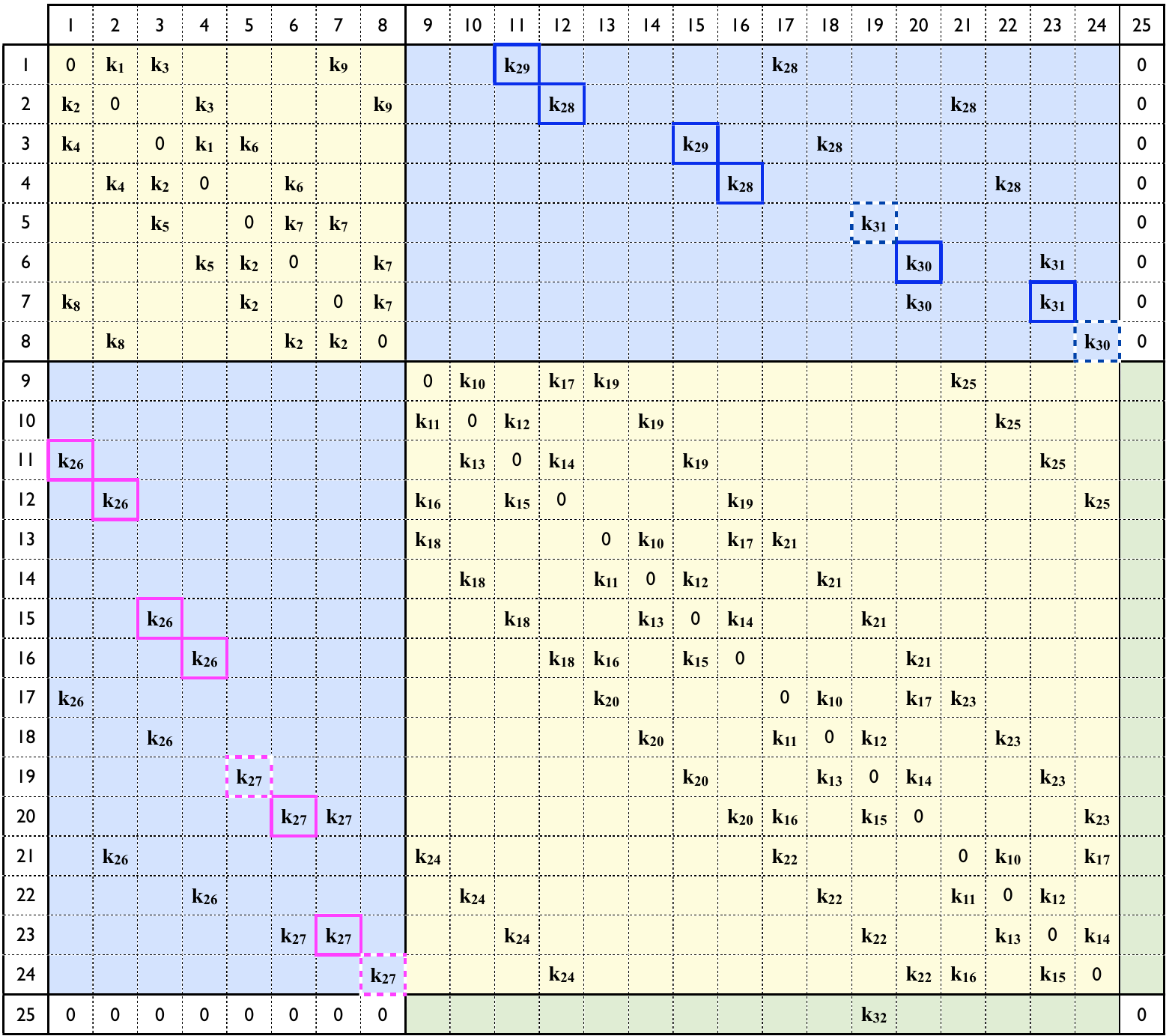}
      \caption{{\bf Matrix of rate constants for the full kinetic model}. The matrix is derived from the allowed transitions pictured in Fig 3. There are a total of 32 unique rate constants. Yellow blocks correspond to chemical transitions within levels I and II, and blue blocks correspond to mechanical transitions between levels I, II, and III. The symmetry seen in the yellow blocks arises from the ``periodic boundary conditions'' within each level. Green blocks correspond to mechanical transitions that lead to detachment of kinesin from the microtubule. Within the blue blocks, cells outlined in magenta represent transitions that take a step backward, and cells outlined in dark blue represent transitions that take a step forward. Cells with dashed outlines correspond to transitions represented by the purple arrows in Fig. 3(b).}
      \label{fig:result_fig5}
   \end{center}
\end{figure}

\clearpage
\begin{figure}
\centering
%\vspace{-0.5cm}
\mbox{\subfigure{\includegraphics[width=2.8in]{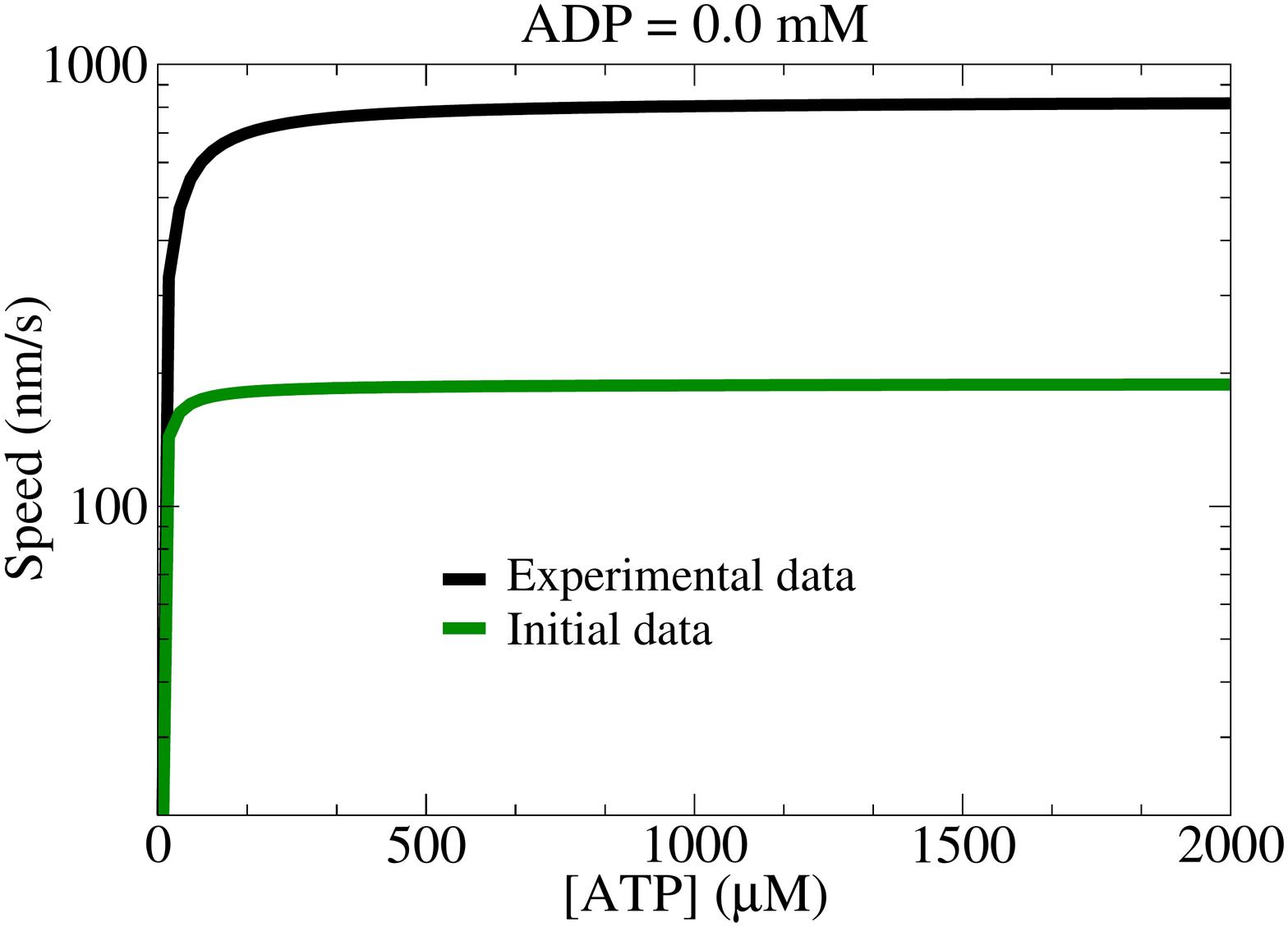}}\quad
\subfigure{\includegraphics[width=2.8in]{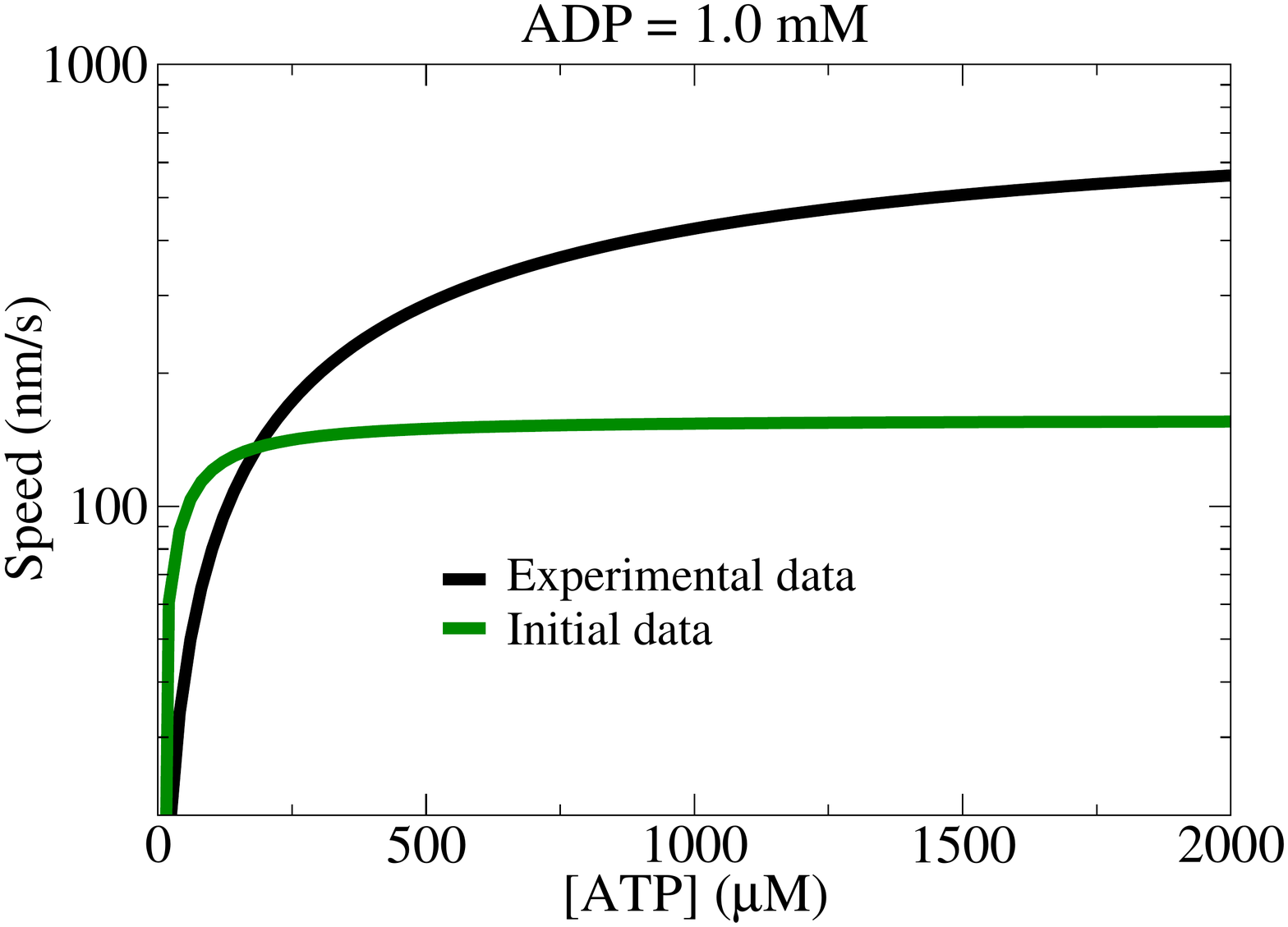} }}
%\end{figure}
%\begin{figure}
%\centering
%\vspace{-5.8cm}
%\mbox{\subfigure{\includegraphics[width=2.8in]{fig4b}}\quad
%\subfigure{\includegraphics[width=2.8in]{fig6b} }}
      \caption{{\bf Initial speed curves.} Black solid lines: Experimental data extracted from \citep{Schief2004,Yajima2002}. Green solid lines: Numerical data obtained from initial rate constants for the model depicted in Fig. 2 via Markov chain calculations. Red dashed lines: Numerical data obtained from optimized rate constants for the model depicted in Fig. 2 via Markov chain calculations.} %Blue points: Numerical data obtained from optimized rate constants via kinetic Monte Carlo.}
      \label{fig:result_fig2}
\end{figure}

%\clearpage
%\begin{figure}
%   \begin{center}
%      \includegraphics[width=5.0in]{blueplot}
%      \caption{{\bf Rate constants from training phase in Experiment B}. Left: The optimized result reproduces experimental data from \citep{Schief2004}  included in objective function, with [ADP] = 1.0 mM and [P$_i$] = 5.0 mM. The blue dashed curve is a fit to the blue experimental points, shown here with experimental error bars. The black points are the calculated results via kinetic Monte Carlo. Right: Average processivity as a function of [ATP], calculated via kinetic Monte Carlo.}
 %     \label{fig:result_fig5}
%   \end{center}
%\end{figure}

\clearpage
\begin{figure}
   \begin{center}
      \includegraphics[width=5.5in]{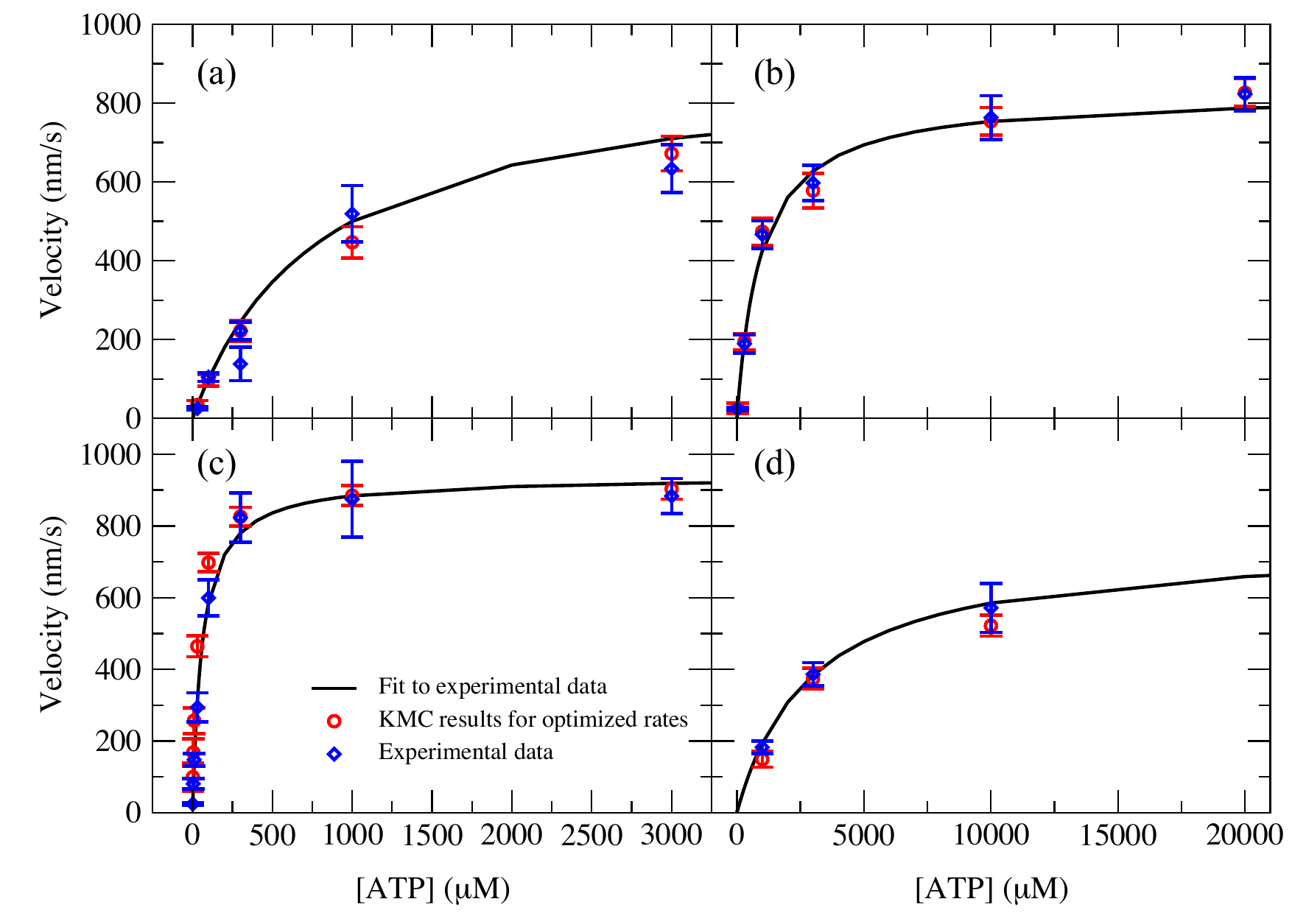}
      \caption{{\bf Rate constants from training phase reproduce Òout-of-sampleÓ experimental results}. The optimized result reproduces experimental data from \citep{Schief2004} included in the optimization scheme (a) and also data that was not previously included in objective function. The training results are shown for (a):  [ADP] = 1.0 mM and [P$_i$] = 5.0 mM. The testing results are for (b):[ADP] = 1.0 mM and [P$_i$] = 0.0 mM; (c): [ADP] = 0.0 mM and [P$_i$] = 10.0 mM; (d): [ADP] = 5.0 mM and [P$_i$] = 0.0 mM. The solid black curves in all plots are fits to the blue experimental points, shown here with experimental error bars. The red points are the calculated results via kinetic Monte Carlo, averaging results from 30 runs, and error bars given by the standard error of the mean.}
      \label{fig:result_fig5}
   \end{center}
\end{figure}

%\clearpage
%\begin{figure}
%   \begin{center}
%\mbox{\subfigure{\includegraphics[width=2.8in]{hist_green}}\quad
%\subfigure{\includegraphics[width=2.8in]{hist_blue} }}
 %     \caption{{\bf Histograms}. Kinetic Monte Carlo allows us to count the number of times a state is visited in each run. Here we depict the average number of times a state (labeled 1 to 25 as in Fig. 2) is visited, for the training set (left) and testing set (right). At high ATP, the 2HB states become more frequent.}
%      \label{fig:result_fig5}
 %  \end{center}
%\end{figure}

\clearpage
\begin{figure}
   \begin{center}
      \includegraphics[width=5.5in]{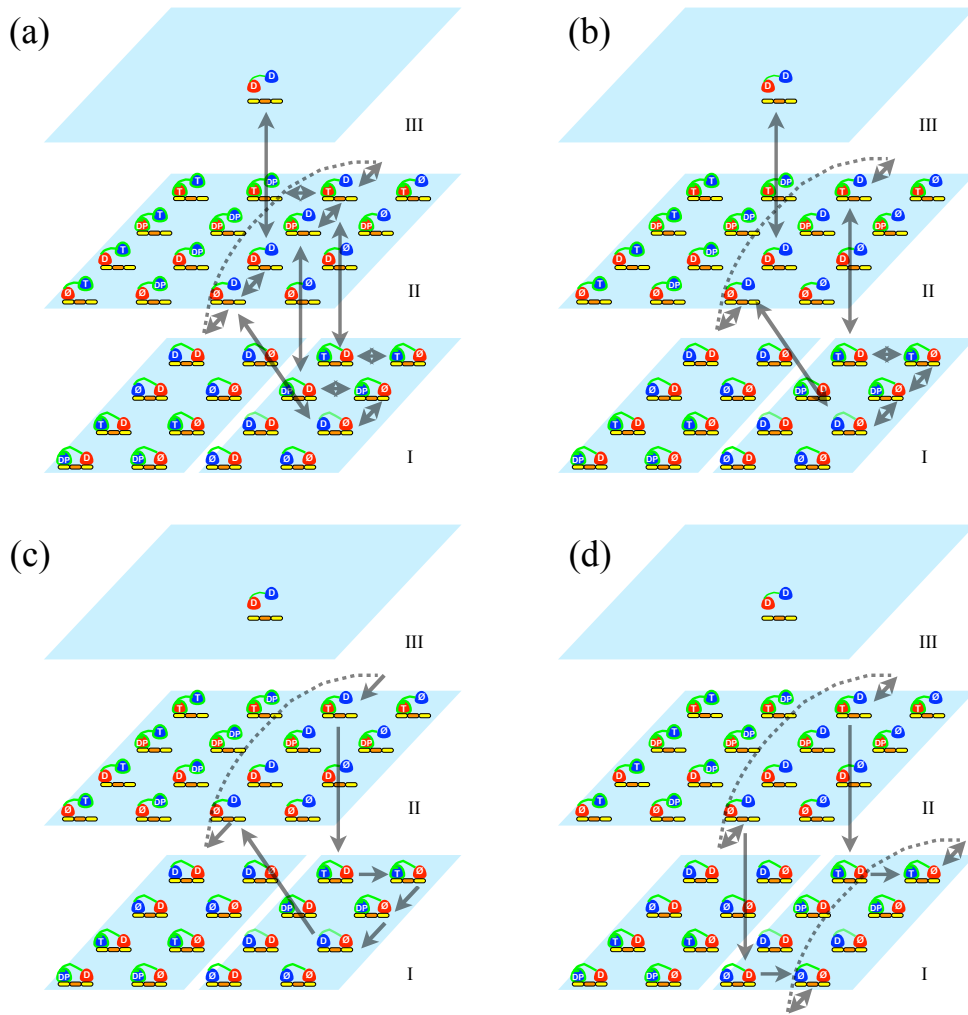}
      \caption{{\bf Pathways proposed in the literature}. Proposed pathways of kinesin walk from previous works are mapped onto our 25-state network. (a) From \citep{Schief2001}, (b) From \citep{Liepelt2010}, (c) From \citep{Hyeon2011}, and (d) From \citep{Clancy2011}. States we have assumed forbidden that nevertheless appear in the aforementioned models were omitted. }
      \label{fig:result_fig10}
   \end{center}
\end{figure}

\clearpage
\begin{figure}
   \begin{center}
      \includegraphics[width=3.5in]{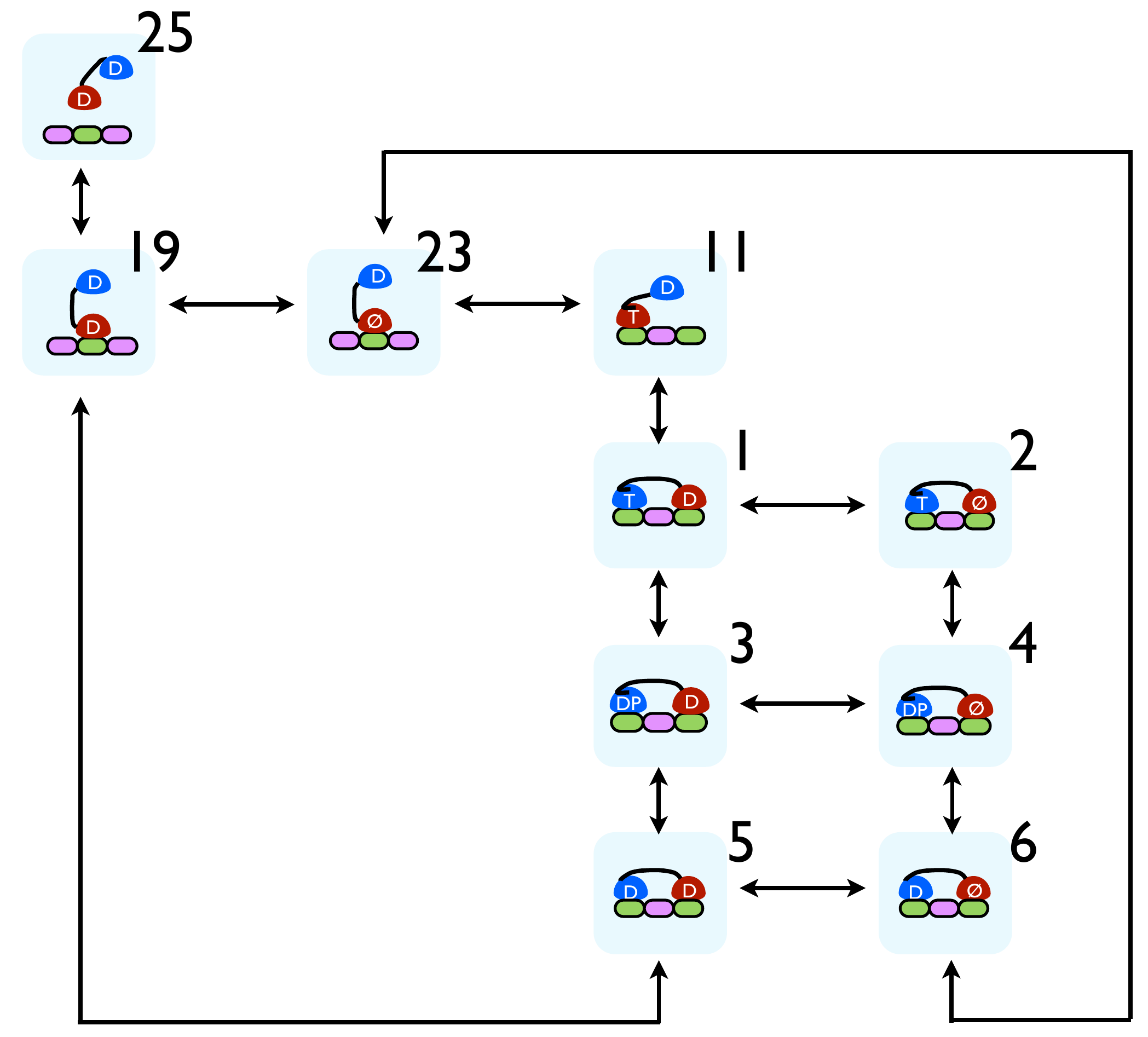}
      \caption{{\bf Dominant pathways}. The optimized rate constants reduce the 25-state network to a small set of dominant cycles visiting more frequently the 10 states shown here. The labels and state representations are the same as in Figure 2.}
      \label{fig:result_fig5}
   \end{center}
\end{figure}

\clearpage
\begin{figure}
   \begin{center}
      \includegraphics[width=3.5in]{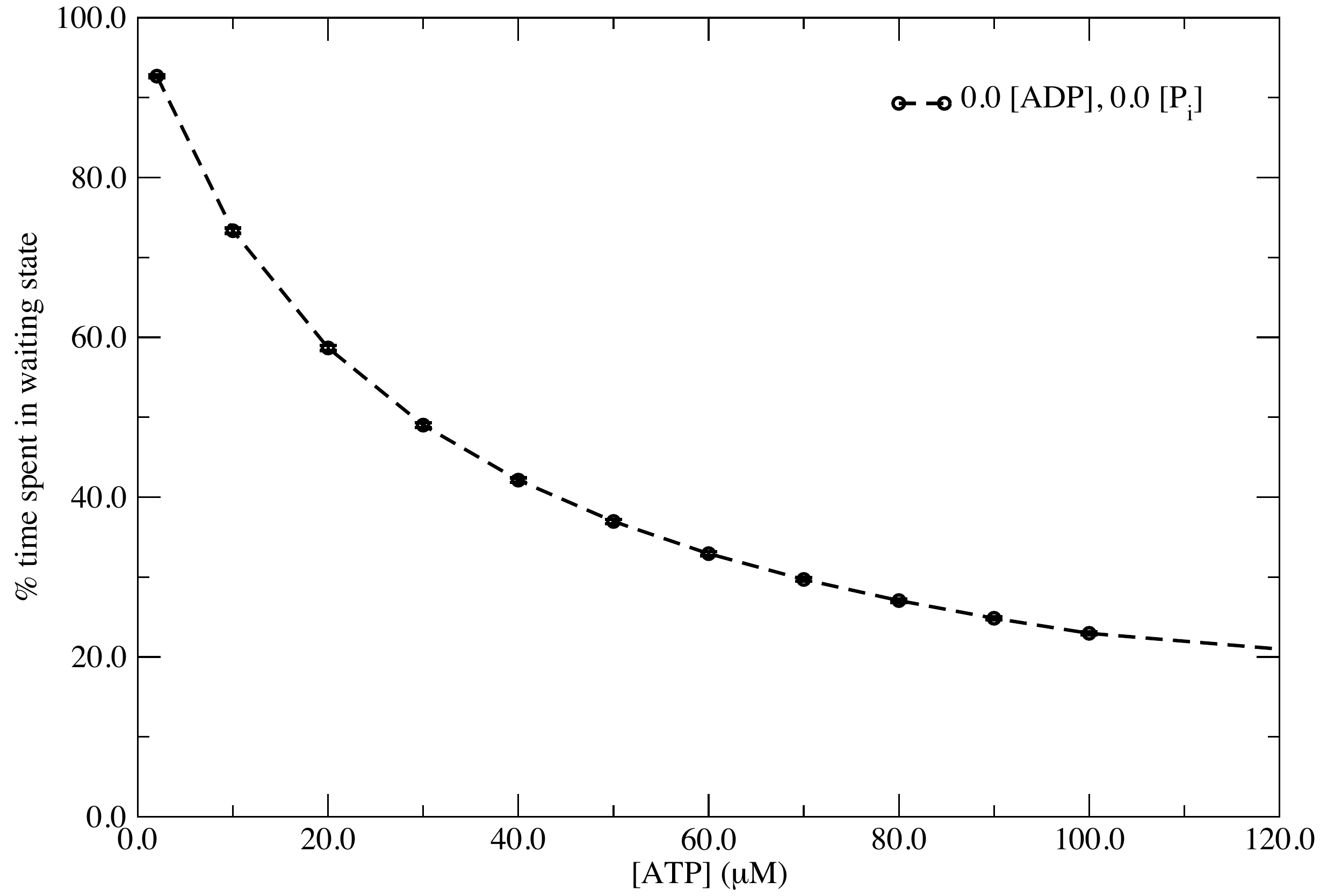}
      \caption{{\bf Time spent in waiting state}. Kinetic Monte Carlo also allows us to find the time spent in each state. At low ATP concentration, the percentage of time spent waiting for ATP binding is $\approx 93\%$, as estimated by Guydosh et al. \citep{Guydosh2009}.}
      \label{fig:result_fig6}
   \end{center}
\end{figure}

% Figures, one per page (fig_1.eps and fig_1.pdf files must be present
% in the document directory)
%\clearpage
%\begin{figure}
%   \begin{center}
%      \includegraphics*[width=3.25in]{fig_1}
%      \caption{}
%      \label{fig:result_fig}
%   \end{center}
%\end{figure}

% closing statement, nothing below matters

\end{document}